\documentclass[prx,reprint,twocolumn,amsmath,superscriptaddress,longbibliography]{revtex4-2}
\usepackage{amsmath,amssymb,bm,mathrsfs,graphicx, braket, times,amsthm,enumerate, scrextend}
\usepackage[colorlinks=true,citecolor=blue,linkcolor=blue]{hyperref}
\usepackage{longtable}
\usepackage{multirow}
\usepackage{array}
\usepackage{bigstrut}
\usepackage[all,cmtip]{xy}
\usepackage[normalem]{ulem}
\usepackage[usenames,dvipsnames]{color}
\usepackage{makecell}
\usepackage{xcolor}
\begin{document}
\title{Catalog of phonon emergent particles and chiral phonons: Symmetry-based classification and materials database investigation}
\author{Houhao Wang}\thanks{The two authors contribute equally to this work.}
\affiliation{National Laboratory of Solid State Microstructures and School of Physics, Nanjing University, Nanjing 210093, China}
\affiliation{Collaborative Innovation Center of Advanced Microstructures, Nanjing University, Nanjing 210093, China}
\author{Dongze Fan}\thanks{The two authors contribute equally to this work.}
\affiliation{National Laboratory of Solid State Microstructures and School of Physics, Nanjing University, Nanjing 210093, China}
\affiliation{Collaborative Innovation Center of Advanced Microstructures, Nanjing University, Nanjing 210093, China}
\affiliation{International Quantum Academy, Shenzhen 518048, China.}
\author{Hoi Chun Po}
\affiliation{Department of Physics, Hong Kong University of Science and Technology, Clear Water Bay, Hong Kong, China}
\author{Xiangang Wan}
\affiliation{National Laboratory of Solid State Microstructures and School of Physics, Nanjing University, Nanjing 210093, China}
\affiliation{Collaborative Innovation Center of Advanced Microstructures, Nanjing University, Nanjing 210093, China}
\affiliation{Hefei National Laboratory, Hefei 230088, China}
\affiliation{Jiangsu Physical Science Research Center, Nanjing, China}
\author{Feng Tang}\email{fengtang@nju.edu.cn}
\affiliation{National Laboratory of Solid State Microstructures and School of Physics, Nanjing University, Nanjing 210093, China}
\affiliation{Collaborative Innovation Center of Advanced Microstructures, Nanjing University, Nanjing 210093, China}

\begin{abstract}
Chirality and topology are fundamental and ubiquitous in nature. 
Symmetry has proven to be a powerful tool for predicting topological phonons.
However, to date, topological phonon emergent particles (EMPs) have not been systematically cataloged in material databases.
Moreover, traditional symmetry methods are often inadequate for predicting chiral phonons, because realistic calculations can yield negative results even when symmetry analysis permits phonon chirality.
Here, we first establish a complete symmetry-based classification: given any space group and Wyckoff positions (WYPOs) occupied by atoms, the number of occurrences of all (co-)irreducible representations ((co-)irreps) (that can host EMPs) can be unambiguously known without omission prior to expensive and parameter-dependent calculation. 
Moreover, whether a phonon mode (belonging to one (co-)irrep) is chiral can also be determined from the occupied WYPOs.  We then perform a materials database investigation identifying over 25 million EMPs at high-symmetry points and along high-symmetry lines and computing the concrete value of phonon angular momentum  for each mode.
We demonstrate two main applications: identifying ideal materials with surface chirality momentum locking and identifying materials with giant phonon magnetic moment.
All computational data are compiled into a website: \href{http://phonon.nju.edu.cn/}{phonon.nju.edu.cn}, which is expected to stimulate future studies on topological and chiral phonons.
\end{abstract}
\maketitle
\date{\today}

\section{Introduction}
\maketitle
Topology and chirality are fundamental concepts in nature: topology describes properties invariant under continuous deformation \cite{topo-book}, while chirality describes a symmetry breaking whereby an object cannot be superimposed on its mirror image \cite{chiral-book}.
Symmetry has been proven to be an powerful tool in the study of topological materials \cite{symmetry1,symmetry2,symmetry3,symmetry4,symmetry5,symmetry6,symmetry7,symmetry8,catalogue1,catalogue2,catalogue3,catalogue4,catalogue5,catalogue6,catalogue7,catalogue8,catalogue9,catalogue10,catalogue11,catalogue12,catalogue13,HPCphonon1,HPCphonon2,HPCphonon3,tangkp1,tangkp2}.
Based on the representation theory of space groups (SGs), Ref. \cite{SciBul-yao} established an exhaustive list of emergent particles (EMPs), including 20 types of spinless particles and 23 types of spinful particles. 
Interestingly, topology and chirality often coexist \cite{TopoChiral1,TopoChiral2,TopoChiral3}, as exemplified by chiral edge states in quantum Hall systems \cite{Hallsystem1,Hallsystem2}, chiral surface Fermi arcs in Weyl semimetals \cite{Weyl-Wan}, and Kramers–Weyl fermions in chiral crystals \cite{Kramers–Weyl-fermions}. This coexistence gives rise to exotic phenomena such as large topological charge \cite{largetopocharge}, long Fermi arcs \cite{longfermiarc1,longfermiarc2}, unusual magnetotransport \cite{unusual-magnetotransport}, a quantized response to circularly polarized light \cite{quantized-circular-photogalvanic-effect}, spin–momentum locking \cite{Spin-momentum-locking1,Spin-momentum-locking2}, and orbital angular momentum (AM) momentum locking \cite{OAM-momentum-locking1}, and can even revolutionize catalytic processes \cite{TopoChiral3}.

On the other hand, for bosonic systems such as phonons, EMPs in the entire frequency window are meaningful and hold potential applications such as phonon waveguides \cite{app-waveguides-1}, abnormal heat transport \cite{app-heat-2}, topological superconductivity \cite{en-super-2}, thermoelectric properties \cite{app-triple}, and catalysis promotion \cite{chemical-1,chemical-2}. 
However, the challenging task of mapping EMPs to specific band crossings hinders the establishment of a materials database.
Meanwhile, chiral phonons that carry finite phonon AM play important roles in the phonon Einstein–de Haas effect \cite{phononHaaseffect}, the spin Seebeck effect \cite{spin-Seebeck-effect}, the thermal Hall effect \cite{thermal-Hall-effect-1,thermal-Hall-effect-2,thermal-Hall-effect-3}, and the phonon magnetic moment (MM) \cite{PMM1,PMM2,PMM3,PMM4}, among other phenomena, and have recently attracted extensive attention \cite{Lifa-L-1,Lifa-L-2,Chiralphonons,chiral-obser,SiO2,Weyl-Chiral,chrialHgS,Chirality,Ueda2025}.
Recently, the coexistence of chiral and topological phonons has become a rapidly developing research frontier \cite{connection1,connection2}, enabling phonon-driven phenomena such as chiral charge density waves \cite{charge-density-wave} and ultrafast demagnetization \cite{ultrafast-demagnetization}, and potentially giving rise to novel quantum states and giant phonon MM.
However, traditional symmetry methods are often inadequate for predicting chiral phonons, for realistic calculations could output negative results through symmetry analysis permits phonon chirality. 
Currently, ideal candidate topological and chiral materials remain scarce, which hinders further experimental and theoretical studies of topological and chiral phonons. Consequently, it is critical to develop a systematic classification of topological and chiral phonons and to build a phonon database to identify ideal materials.

\begin{figure*}[!t]
	\centering\includegraphics[width=1\textwidth]{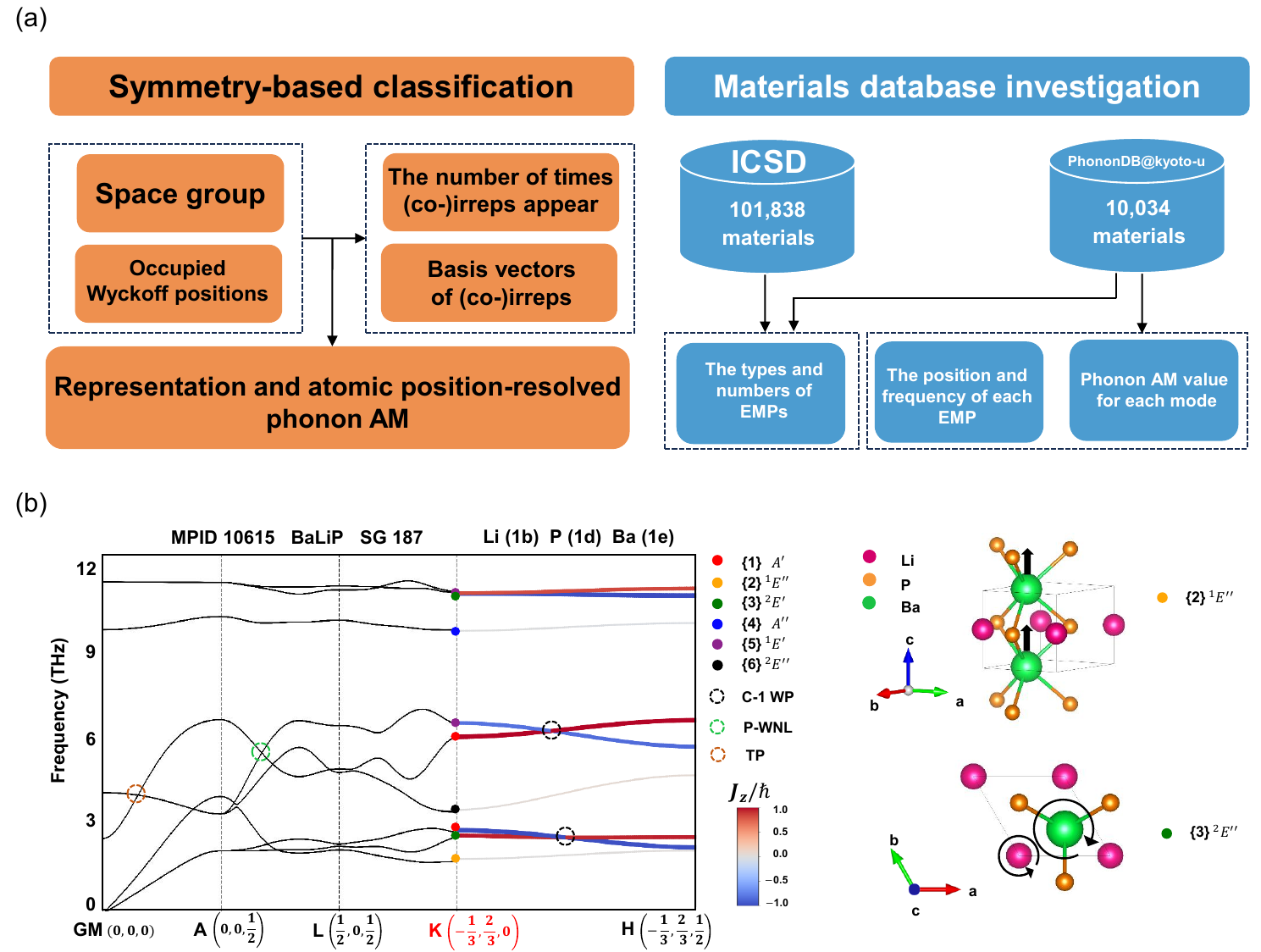}
	\caption{
Schematic workflow of this work (a) and an illustrative material example (b). (a) Symmetry-based classification: Given any SG and the occupied WYPOs, the multiplicities of (co-)irreps and expressions for the phonon AM can be determined. Materials database investigation: Using the symmetry results, we qualitatively identify EMPs at HSPs for each material in ICSD and PhononDB@Kyoto-u. For PhononDB@Kyoto-u, we further quantify the EMP frequencies at HSPs and along HSLs, and calculate the phonon AM for each mode. (b) Prototypical material example (BaLiP, MPID 10615, SG 187): There are three atoms per unit cell, with Li, P, and Ba occupying WYPOs 1b, 1d, and 1e, respectively. Based on the symmetry results in Table~\ref{tab}, the co-irreps $\{1\}$, $\{3\}$, and $\{5\}$ appear twice at the K point, whereas co-irreps $\{2\}$, $\{4\}$, and $\{6\}$ appear once. Phonon modes belonging to co-irreps $\{1\}$, $\{3\}$, and $\{5\}$ carry phonon AM along the $z$ direction, while those belonging to co-irreps $\{2\}$, $\{4\}$, and $\{6\}$ carry zero phonon AM. For instance, co-irrep $\{3\}$ corresponds to a chiral mode with Li and Ba rotating oppositely in the $xy$ plane, whereas co-irrep $\{2\}$ corresponds to a linearly polarized mode (Ba vibrating along $z$). Moreover, symmetry results indicate that all phonon modes along the K$-$H path can carry nonzero phonon AM along the $z$ direction. $J_z$ represents the $z$ component of the phonon AM along K$-$H. Note that two accidental C-1 WPs with $J_z$ values of $-0.057\hbar$ and $0.165\hbar$ arise from crossings between bands with opposite signs of phonon AM along K$-$H. Paths GM$-$A and A$-$L host P-WNL and TP, respectively.
 }
	\label{Flowchart}
\end{figure*}

\begin{table*}[!t]
\caption{
Symmetry results at the K $\left(-\frac{1}{3}, \frac{2}{3}, 0\right)$ point of SG 187.
The six one-dimensional co-irreps in the first column correspond to $\mathrm{A}^\prime$, $^{1}\mathrm{E}^{\prime\prime}$, $^{2}\mathrm{E}^\prime$, $\mathrm{A}^{\prime\prime}$, $^{1}\mathrm{E}^\prime$, and $^{2}\mathrm{E}^{\prime\prime}$ in the Mulliken notation \cite{Mulliken1939}, respectively.
For each co-irrep, the second column gives the symmetry-allowed phonon AM expression, $\left(0,0,r_1\right)$, where $r_1$ is a real coefficient.
The third column lists, for a single atom placed at a given WYPO, the multiplicity of the co-irrep induced by that WYPO (numbers in parentheses).
For example, when WYPO 1b is occupied once, co-irreps \{3\}, \{4\}, and \{5\} will appear once, respectively.
For a specific material, the total multiplicity is obtained by summing the contributions from its occupied WYPOs.
WYPOs that allow a nonvanishing phonon AM (i.e., $r_1\neq 0$) are printed in bold.
For example, a phonon mode corresponding to co-irrep \{2\} can carry nonvanishing phonon AM when the atoms occupy at least one of the WYPOs \{2g, 2h, 6n, and 12o\}.
}\label{tab}
   \begin{tabular}{m{1.5cm}<{\centering}m{3cm}<{\centering}p{12.9cm}}
\hline
\hline

\noalign{\vskip 4pt}
\multicolumn{3}{c}{$\text{K}$ $\left(-\frac{1}{3},\frac{2}{3},0\right)$ of SG No. 187}\\ 
\noalign{\vskip 4pt}
\hline
\noalign{\vskip 3pt}
\multirow{1}{*}{co-irrep} & \multicolumn{1}{c}{Phonon AM} & \multicolumn{1}{c}{WYPO information} \\  
\noalign{\vskip 3pt} 
\hline

\noalign{{\vskip 3pt}}
\{1\} & $\left(0,0,r_1\right)$ & 1a: (0); \enspace 1b: (0); \enspace \textbf{1c: (1)}; \enspace \textbf{1d: (1)}; \enspace \textbf{1e: (1)}; \enspace \textbf{1f: (1)}; \enspace 2g: (1); \enspace \textbf{2h: (1)}; \enspace \textbf{2i: (1)}; \enspace \textbf{3j: (2)}; \enspace \textbf{3k: (2)}; \enspace \textbf{6l: (4)}; \enspace \textbf{6m: (4)}; \enspace \textbf{6n: (3)}; \enspace \textbf{12o: (6)}\\
\noalign{{\vskip 3pt}}
\{2\} & $\left(0,0,r_1\right)$ & 1a: (0); \enspace 1b: (0); \enspace 1c: (0); \enspace 1d: (0); \enspace 1e: (1); \enspace 1f: (1); \enspace \textbf{2g: (1)}; \enspace \textbf{2h: (1)}; \enspace 2i: (1); \enspace 3j: (1); \enspace 3k: (1); \enspace 6l: (2); \enspace 6m: (2); \enspace \textbf{6n: (3)}; \enspace \textbf{12o: (6)}\\
\noalign{{\vskip 3pt}}
\{3\} & $\left(0,0,r_1\right)$ & \textbf{1a: (1)}; \enspace \textbf{1b: (1)}; \enspace 1c: (0); \enspace 1d: (0); \enspace \textbf{1e: (1)}; \enspace \textbf{1f: (1)}; \enspace \textbf{2g: (1)}; \enspace 2h: (1); \enspace \textbf{2i: (1)}; \enspace \textbf{3j: (2)}; \enspace \textbf{3k: (2)}; \enspace \textbf{6l: (4)}; \enspace \textbf{6m: (4)}; \enspace \textbf{6n: (3)}; \enspace \textbf{12o: (6)}\\
\noalign{{\vskip 3pt}}
\{4\} & $\left(0,0,r_1\right)$ & 1a: (1); \enspace 1b: (1); \enspace 1c: (0); \enspace 1d: (0); \enspace 1e: (0); \enspace 1f: (0); \enspace 2g: (1); \enspace \textbf{2h: (1)}; \enspace \textbf{2i: (1)}; \enspace 3j: (1); \enspace 3k: (1); \enspace 6l: (2); \enspace 6m: (2); \enspace \textbf{6n: (3)}; \enspace \textbf{12o: (6)}\\
\noalign{{\vskip 3pt}}
\{5\} & $\left(0,0,r_1\right)$ & \textbf{1a: (1)}; \enspace \textbf{1b: (1)}; \enspace \textbf{1c: (1)}; \enspace \textbf{1d: (1)}; \enspace 1e: (0); \enspace 1f: (0); \enspace \textbf{2g: (1)}; \enspace \textbf{2h: (1)}; \enspace 2i: (1); \enspace \textbf{3j: (2)}; \enspace \textbf{3k: (2)}; \enspace \textbf{6l: (4)}; \enspace \textbf{6m: (4)}; \enspace \textbf{6n: (3)}; \enspace \textbf{12o: (6)}\\
\noalign{{\vskip 3pt}}
\{6\} & $\left(0,0,r_1\right)$ & 1a: (0); \enspace 1b: (0); \enspace 1c: (1); \enspace 1d: (1); \enspace 1e: (0); \enspace 1f: (0); \enspace \textbf{2g: (1)}; \enspace 2h: (1); \enspace \textbf{2i: (1)}; \enspace 3j: (1); \enspace 3k: (1); \enspace 6l: (2); \enspace 6m: (2); \enspace \textbf{6n: (3)}; \enspace \textbf{12o: (6)}\\
\noalign{{\vskip 3pt}}
\hline
\hline
  \end{tabular}
\end{table*}

In this paper, to overcome these challenges, we first establish a complete symmetry-based classification (Fig. \ref{Flowchart}(a)): given any SG and occupied Wyckoff positions (WYPOs) by atoms, the number of occurrences of all irreducible (co-)representations ((co-)irreps) can be fully determined. Moreover, the phonon AM for each phonon mode (belonging to one (co-)irrep) and the WYPOs from which phonon AM originates can also be determined.
Based on our symmetry results, the number of EMPs can be qualitatively determined without explicit calculation.
Phonon AM behaviors can be illustrated from the perspective of occupied WYPOs and basis vectors of (co-)irreps.
We apply our symmetry results to perform a database investigation by qualitatively identifying over 25 million EMPs and quantitatively calculating their frequencies at HSPs and along HSLs, and by computing phonon AM for each phonon mode, with all computational data for over 111,872 crystalline compounds compiled into an online database.
We demonstrate the application of the database to search for ideal materials with chirality momentum locking, giant phonon MM, coexisting EMPs, petal-shaped surface states induced by nearly nodal lines, and so forth.
Our work can resolve phonon AM and provides a powerful guide for searching or designing ideal EMPs and chiral phonons, and the database will catalyze the discovery of new candidates.

\section{Results}
\subsection{Symmetry-based results and illustrative example}
We compile the symmetry-based classification results into user-friendly tabulations, systematically covering the HSPs, HSLs, and high-symmetry planes (HSPLs) of all 230 SGs. All results are provided in Sec. 4 of the Supplementary Material (SM).
These tabulations serve as a powerful guide for searching and designing materials that host both EMPs and phonon AM.
Here, we demonstrate the use of symmetry results at the K $\left(-\frac{1}{3}, \frac{2}{3}, 0\right)$ point of SG 187, as presented in Table \ref{tab}.
When WYPO 1b is occupied once, co-irreps \{3\}, \{4\}, and \{5\} will appear once, respectively. The phonon mode corresponding to co-irrep \{2\} can carry a nonvanishing phonon AM of the form $\left(0,0,r_1\right)$, where $r_1$ is a nonzero real coefficient, when the atoms occupy at least one of the WYPOs \{2g, 2h, 6n, and 12o\}.

Here, based on Table \ref{tab}, we showcase a material example illustrating the phonon AM behaviors from the perspective of occupied WYPOs (Fig. \ref{Flowchart}(b)).
BaLiP with Materials Project Identification number (MPID) 10615 and SG 187 has three atoms per unit cell, with Li, P, and Ba atoms occupying WYPOs 1b, 1d, and 1e respectively. 
For co-irrep \{3\}, WYPO 1b (Li atom) and 1e (Ba atom) each contribute once, both contributing nonzero phonon AM along the $z$ direction. The two occurrences of \{3\} are at the 2nd and 8th energy levels at the K point, with phonon AM values of $(0, 0, \pm0.970\hbar)$ ($\hbar$ is the reduced Planck constant), respectively. In real space, these two modes correspond to circular polarizations: The Li atom and Ba atom perform circular motion along the eigenvectors of a circularly polarized optical phonon mode in the $xy$ plane.
A similar analysis reveals that co-irrep \{1\} and \{5\} each occur twice, with the two occurrences of \{1\} at the 3rd and 5th energy levels at the K point having phonon AM values of $(0, 0, \mp 0.999\hbar)$, and the two occurrences of \{5\} at the 6th and 9th energy levels at the K point having phonon AM values of $(0, 0, \mp 0.817\hbar)$.
For co-irrep \{2\}, only WYPO 1e (Ba atom) contributes once but does not contribute phonon AM. The single occurrence of \{2\} is at the 1st energy level at the K point with no phonon AM. The Ba atom performs a linearly polarized vibration along the $z$ direction in real space.
A similar analysis shows that \{2\} and \{4\} each occur once at the K point, at the 7th and 4th energy levels, respectively, and both carry zero phonon AM.

Furthermore, based on the symmetry results along the K--H path (see Sec.~4 of the SM), we can know that all bands along the K--H path carry phonon AM along the \(z\) direction.
The bands and phonon AM along the K$-$H path are shown in Fig. \ref{Flowchart}b. Two accidental C-1 WPs, with phonon AM values of $(0,0,-0.057\hbar)$ and $(0,0,0.165\hbar)$, arise from the crossing of bands with opposite phonon AM. Additionally, the GM$-$A and A$-$L paths also host P-WNL and TP, respectively. So accidental EMPs with nonzero phonon AM along HSLs provide an excellent platform to study the connection between topology and chirality. 

\begin{figure*}[!t]
	\centering\includegraphics[width=1\textwidth]{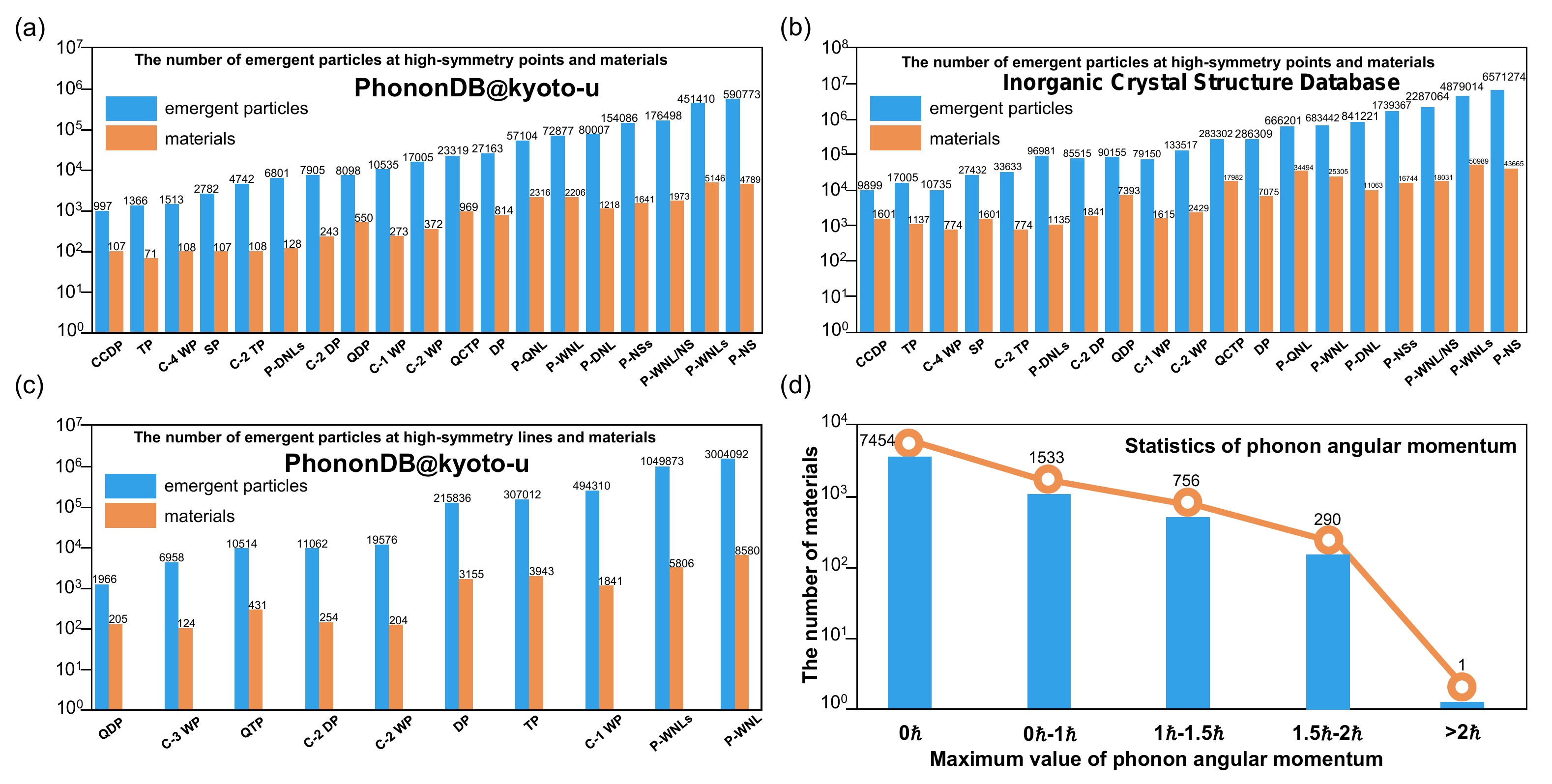}
	\caption{Statistical overview of the materials database investigation. (a-c) Statistics of EMPs identified in materials from PhononDB@Kyoto-u and ICSD, and the number of materials hosting each EMP type. The blue bars represent the number of EMPs. The orange bars represent the number of materials. The numbers on the bars indicate the exact counts. Panels (a) and (b) summarize all 19 symmetry-allowed EMP types at HSPs based on PhononDB@Kyoto-u and ICSD. Panel (c) summarizes all 10 accidental EMP types along HSLs based on PhononDB@Kyoto-u. The full names of EMPs are from Ref. \cite{SciBul-yao}. (d) Distribution of the maximum absolute value of phonon AM for 10,034 materials in PhononDB@Kyoto-u.}
	\label{Statistics}
\end{figure*}

\begin{figure*}[!hbtp]
	\centering\includegraphics[width=1\textwidth]{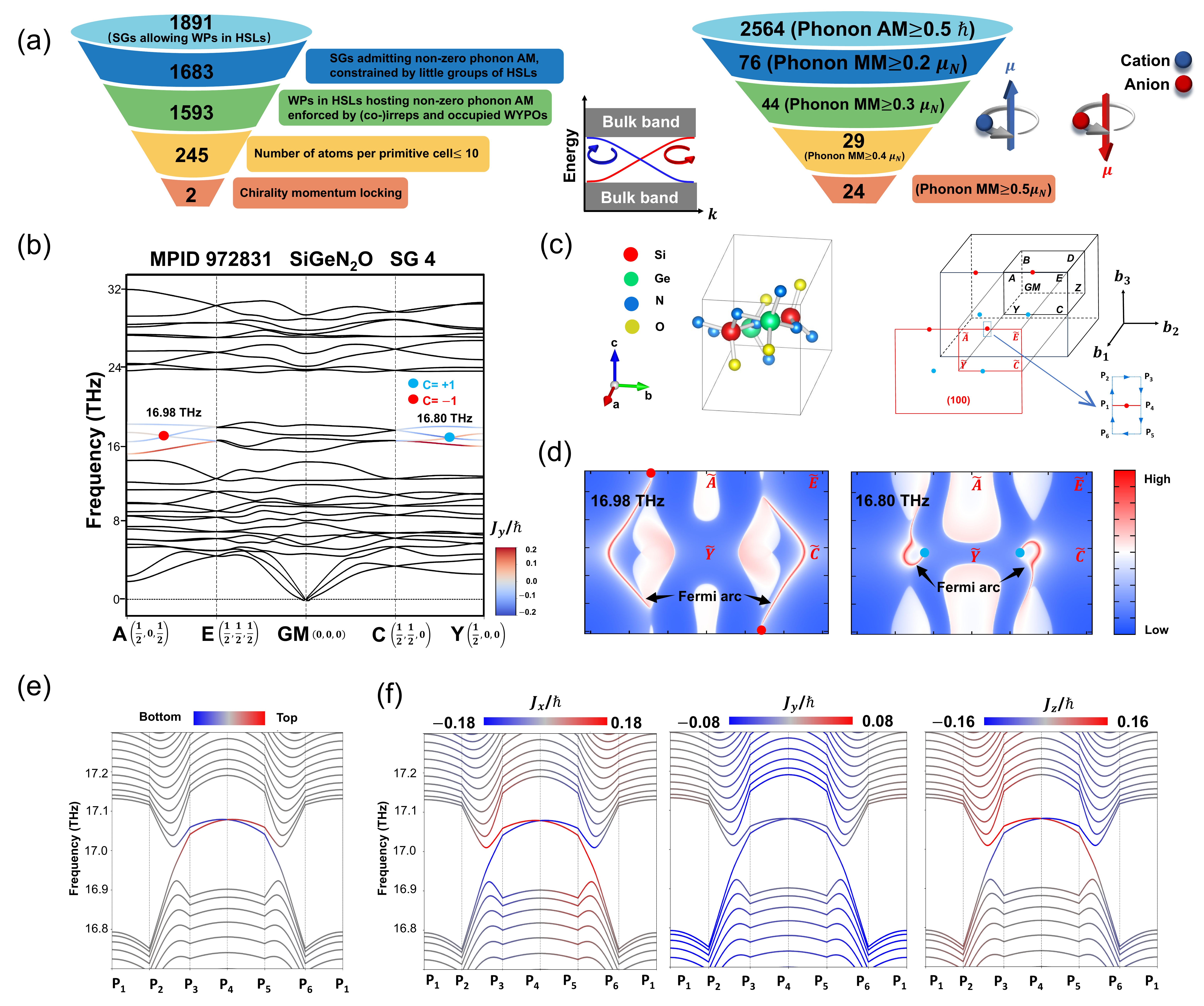}
\caption{
(a)  Workflow for material screening of chirality momentum locking (left) and giant phonon MM in phonon modes (right). The numbers in the workflows represent the number of materials at each stage of the screening process.
(b-f) Material example (SiFeN$_2$O, MPID 972831, SG 4) with chirality momentum locking.
(b) Phonon spectrum, WPs, and phonon AM. Phonon modes along paths A$-$E and C$-$Y can carry nonzero phonon AM along the $y$ direction ($J_y$ represents the $y$ component of the phonon AM) based on symmetry results. Note that there is one WP with charge $-1$ ($J_y=-0.07\hbar$) at 16.98 THz along A$-$E, and one with charge $+1$ ($J_y=-0.06\hbar$) at 16.80 THz along C$-$Y, formed by the 18th and 19th bands, respectively.
(c) Crystal structure, bulk BZ, and (100) surface BZ. The bulk WPs and the projected four WPs formed by the 18th and 19th bands are denoted.
(d) Isofrequency surface contours on the (100) surface at 16.98 THz and 16.80 THz, respectively. The clear Fermi arcs originating from the WPs in (c) are denoted.
(e) Slab bands along a closed blue contour around one of the WPs in (c), revealing two unidirectional surface states. (f) $x, y, z$ component of the phonon AM of Slab bands in (e).
The $y$ component relative to the $x$ and $z$ components of Phonon AM can be neglected. Chirality momentum locking: Left-moving and right-moving surface states have opposite phonon AM along the $x$ and $z$ directions.
}
\label{Application}
\end{figure*}

\begin{table*}
\caption{The list of materials with giant phonon MM exceeding $0.5\,\mu_N$ includes the SG, chemical formula, MPID, the maximum phonon MM, and its location in the BZ.
Note that some phonon modes hosting giant phonon MM also correspond to C-1 WP or L-NS. Here, $\mu_N=e\hbar/(2m_p)$ is the nuclear magneton, where $m_p$, $e$, and $\hbar$ denote the proton rest mass, elementary charge, and reduced Planck constant, respectively.
}\label{tabPMM}
\begin{tabular}{m{1cm}<{\centering}p{17cm}}
\hline
\hline

\noalign{\vskip 3pt}
\multirow{1}{*}{SG} & \multicolumn{1}{c}{Material with giant phonon MM} \\  
\noalign{\vskip 3pt} 
\hline

\noalign{\vskip 3pt} 
8&\multicolumn{1}{c}{U $\left( - \frac{v}{2}, \frac{v}{2},  \frac{1}{2}\right)$}\\
\noalign{\vskip 3pt} 
&\text{H$_{3}$BrO (MPID 625521): 0.62 $\mu_N$}\\
\noalign{\vskip 3pt}
\hline 

\noalign{\vskip 3pt} 
19&\multicolumn{1}{c}{D $\left( \frac{1}{2}, v, 0\right)$}\\
\noalign{\vskip 3pt} 
&\text{HBr (MPID 32684): 0.508 $\mu_N$ (L-NS)}\\
\noalign{\vskip 3pt}
\hline 

\noalign{\vskip 3pt} 
31&\multicolumn{1}{c}{C $\left( u, \frac{1}{2}, 0\right)$}\\
\noalign{\vskip 3pt} 
& \text{GaHO$_{2}$ (MPID 23803): 0.564 $\mu_N$} \enspace \text{AlHO$_{2}$ (MPID 23807): 0.637 $\mu_N$}\\
\noalign{\vskip 3pt}
\hline 

\noalign{\vskip 3pt} 
36&\multicolumn{1}{c}{A $\left( \frac{u}{2}, \frac{u}{2}, \frac{1}{2}\right)$}\\
\noalign{\vskip 3pt} 
&\text{AlHO$_{2}$ (MPID 625056): 0.716 $\mu_N$ (L-NS)} \enspace \text{AlHO$_{2}$ (MPID 626902): 0.716 $\mu_N$ (L-NS)}  \enspace \text{AlHO$_{2}$ (MPID 625054): 0.716 $\mu_N$ (L-NS)} \enspace \text{CsHO (MPID 625136): 0.987 $\mu_N$ (L-NS)}\\
\noalign{\vskip 3pt}
&\multicolumn{1}{c}{D $\left( 0, \frac{1}{2}, w\right)$}\\
\noalign{\vskip 3pt}
&\text{ScHO$_{2}$ (MPID 625150): 0.646 $\mu_N$} \enspace \text{RbHO (MPID 626620): 0.931 $\mu_N$ (C-1 WP)} \enspace \text{KHO (MPID 626785): 0.996 $\mu_N$ (C-1 WP)}\\
\noalign{\vskip 3pt} 

\noalign{\vskip 3pt}
&\multicolumn{1}{c}{SM $\left( \frac{u}{2},  \frac{u}{2}, 0\right)$}\\
&\text{HBr (MPID 632229): 1.27 $\mu_N$}\\
\noalign{\vskip 3pt} 

\noalign{\vskip 3pt}
\hline

\noalign{\vskip 3pt} 
44&\multicolumn{1}{c}{Q $\left( \frac{v}{2}, \frac{1}{2} - \frac{v}{2}, \frac{v}{2}\right)$}\\
\noalign{\vskip 3pt} 
&\text{H$_{4}$IN (MPID 34381): 0.659 $\mu_N$}\\
\noalign{\vskip 3pt}
\hline

\noalign{\vskip 3pt} 
152&\multicolumn{1}{c}{DT $\left( u, 0, 0\right)$}\\
\noalign{\vskip 3pt} 
&\text{WN$_{2}$ (MPID 754629): 0.552 $\mu_N$ (C-1 WP)}\\
\noalign{\vskip 3pt}
\hline

\noalign{\vskip 3pt} 
156&\multicolumn{1}{c}{K $\left( -\frac{1}{3}, \frac{2}{3}, \frac{1}{2}\right)$}\\
&\text{SrGaSnH (MPID 978852): 1.763 $\mu_N$}\\
\noalign{\vskip 3pt}
&\multicolumn{1}{c}{H $\left( -\frac{1}{3}, \frac{2}{3}, 0\right)$}\\
\noalign{\vskip 3pt} 
&\text{BaAlSiH (MPID 571093): 1.553 $\mu_N$} \enspace \text{SrAlSiH (MPID 570485): 1.636 $\mu_N$} \enspace \text{SrAlGeH (MPID 980057): 1.684 $\mu_N$} \enspace \text{SrGaSiH (MPID 979137): 1.835 $\mu_N$}\\
\noalign{\vskip 3pt} 
\hline

\noalign{\vskip 3pt} 
160&\multicolumn{1}{c}{Y $\left( u + \frac{1}{2}, \frac{1}{2}, \frac{1}{2} - u\right)$}\\
\noalign{\vskip 3pt} 
&\text{H$_{3}$BrO (MPID 625509): 0.648 $\mu_N$}\\
\noalign{\vskip 3pt}
\hline

\noalign{\vskip 3pt} 
187&\multicolumn{1}{c}{H $\left( -\frac{1}{3}, \frac{2}{3}, 0\right)$}\\
\noalign{\vskip 3pt} 
&\text{HoHSe (MPID 24012): 0.627 $\mu_N$} \enspace \text{YHSe (MPID 28797): 0.631 $\mu_N$}\\
\noalign{\vskip 3pt}
\hline

\noalign{\vskip 3pt} 
215&\multicolumn{1}{c}{SM $\left( u, u,  0\right)$}\\
\noalign{\vskip 3pt} 
&\text{H$_{4}$NCl (MPID 34337): 0.542 $\mu_N$} \enspace \text{H$_{4}$BrN (MPID 36248): 0.591 $\mu_N$}\\
\noalign{\vskip 3pt}

\hline
\hline
\end{tabular}
\end{table*}

\subsection{Database investigation results and statistics}
We apply our symmetry results to Inorganic Crystal Structure Database (ICSD) \cite{ICSD} with 101,838 materials and the Phonon Database at Kyoto University (PhononDB@kyoto-u) \cite{phonopyDB} with 10,034 materials to perform materials database investigations at HSPs and along HSLs based on the workflow (right panel of Fig. \ref{Flowchart}(a)). 
We qualitatively identify 20,516,167 EMPs at HSPs, with over 90\% of the materials hosting such EMPs. The statistics for EMPs and their hosting materials at HSPs are presented in Figs. \ref{Statistics}(a) and \ref{Statistics}(b).
Furthermore, for PhononDB@kyoto-u, the frequencies of all EMPs at HSPs are also quantitatively determined by the force constants, and the phonon AM value for each mode are also calculated.
We further identify the positions and frequencies of over 5 million accidental EMPs along HSLs across the entire frequency window for each material in PhononDB@kyoto-u, and the phonon AM value for each mode are also calculated. 
The statistics for accidental EMPs and their hosting materials along HSLs are presented in Fig. \ref{Statistics}(c).
The distribution of the maximum absolute value of phonon AM for all materials in PhononDB@kyoto-u is shown in Fig.~\ref{Statistics}(d).
The largest phonon AM is $(-2.249\hbar, 0, 0)$, which originates from an accidental Dirac point along the HSL P $\left(\frac{1}{2}, v, \frac{1}{2}\right)$ in ZnAgPS$_4$ (MPID 558807, SG 33).
Our symmetry results are both consistent with the quantitative calculations of EMP numbers and effective in predicting the phonon AM.

All computational data are integrated into a website \href{http://phonon.nju.edu.cn/}{phonon.nju.edu.cn}.
Users can search for specific EMPs, chemical compositions, crystal symmetries, ICSD numbers, or MPIDs to identify relevant compounds for further study.
Additionally, we provide the character tables, along with the coordinates of the WYPOs in 230 SGs, the $k \cdot p$ effective models for each (co-)irrep, as well as information such as the EMP types and the nearly nodal structures corresponding to (co-)irreps used in our study.
The details of usage of our database are provided in the Sec. S2.2 of the SM. 
The SM also includes list of 2,309 nearly ideal topological phonon materials (Sec. S2.3 of the SM) and 559 layered materials (Sec. S2.4 of the SM) in our database, a list of symmetry-enforced EMPs at HSPs by (co-)irreps and WYPOs (Sec. S3 of the SM), as well as the atomic insulator basis sets for 230 SGs (Sec. S5 of the SM).

\section{Application}
Our database can be used to investigate topological properties, chirality, and their coexistence, and it offers multiple intriguing applications. 

We demonstrate how researchers can use the online database to efficiently screen materials with coexisting EMPs and identify promising candidates (Sec. S2.5a of the SM).
Nearly nodal structures can lead to significant physical consequences, such as large Berry curvature and quasi-quantized Berry phases \cite{tangkp2, NNL2, NNL3, NNL4, NNL5, NNL6, NNL7}.
We first identify petal-shaped surface states induced by four nearly straight nodal lines emanating from the C-4 WP in the ideal material PtSiSr (SG 198, ICSD number 604437) (Sec. S2.5b of the SM). We also show how the online database can be used to obtain phonon pseudoangular momentum in chiral materials (Sec. S2.5c of the SM). 
Finally, we present two representative applications: chirality momentum locking and giant phonon MM, in Sec.~\ref{CML} and Sec.~\ref{PMM}, respectively.

\subsection{Chirality momentum locking of phonon surface state}\label{CML}
Momentum space textures of phonon AM for bulk phonon states have been investigated \cite{phononAMtexture1,phononAMtexture2,phononAMtexture3,phononAMtexture4}.
However, to our knowledge, phonon AM for phonon surface states remains unexplored.
WPs located in HSLs provide an ideal platform for investigating phonon AM for surface states: on the one hand, they can induce unidirectional edge states; on the other hand, based on our symmetry-based classification results, they can often carry nonzero phonon AM. 
Is it possible that the phonon AM for surface states is highly correlated with the momentum direction (shown in the chiral momentum locking schematic in Fig. \ref{Application}(a))?

Here, based on the workflow shown in the left panel of Fig.~\ref{Application}(a), we select 245 materials hosting WPs with nonzero phonon AM on HSLs, each containing no more than 10 atoms per primitive cell.
We manually selected two high-quality materials exhibiting chirality momentum locking: SiGeN$_2$O (MPID 972831, SG 4) and TeO$_2$ (MPID 8377, SG 19). In both materials, all WPs are located on HSLs and are well separated in energy from other bands. Moreover, the Fermi arcs and the unidirectional surface states associated with these WPs are exceptionally clean, thereby facilitating experimental observation.
Here, we use SiGeN$_2$O with MPID 972831 and SG 4 as an example (Fig. \ref{Application}(c)). The other material is provided in Sec. S2.5C of the SM.

The 18th and 19th bands form four C-1 WPs in the bulk Brillouin zone (BZ), all of which appear along HSLs. As shown in the phonon spectra of Fig. \ref{Application}(b), one C-1 WP with charge $-1$ and phonon AM $(0, -0.07\hbar, 0)$ appears along the A$-$E path, and another with charge $-1$ and phonon AM $(0, -0.06\hbar, 0)$ appears along the C$-$Y path. In addition, time-reversal symmetry results in the existence of two additional WPs on the opposite paths.
The bulk BZ and (100) surface BZ with projected these two pair WPs are presented in Fig. \ref{Application}(c).
Fig. \ref{Application}(d) shows the corresponding clear Fermi arc states from these two pair WPs. We select a closed path with vertices located at one of the projected WPs on (100) surface (shown as a blue rectangle in Fig. \ref{Application}(c)) to calculate slab energy bands. The corresponding surface states are presented in Fig. \ref{Application}(e), which exhibit a right-moving edge mode on the top surface and a left-moving chiral edge mode on the bottom surface, emerging inside the band gap.
We then calculate the phonon AM of the slab bands using their wavefunctions (Fig.~\ref{Application}(f)).
The $x$ and $z$ components dominate, while the $y$ component is negligible.
The two unidirectional edge modes have opposite phonon AM components along the $x$ and $z$ directions (Fig.~\ref{Application}(f)).
As a consequence of chirality momentum locking, phonon backscattering is suppressed, leading to robust unidirectional phonon surface states with potential applications in thermal diodes, thermal transistors, and other thermal devices.

\subsection{Giant phonon magnetic moment}\label{PMM}
The phonon AM $\bm{L}_m$ of a charged ion labeled by $m$ induces a phonon MM $\bm{\mu}_m$ through the gyromagnetic-ratio tensor $\bm{\gamma}_m$, such that $\bm{\mu}_m=\bm{\gamma}_m\bm{L}_m$ \cite{orbitalMM1,orbitalMM2,SiO2}.
A large phonon AM together with a large gyromagnetic ratio has the potential to generate a giant phonon MM. The total phonon MM is given by $\bm{\mu}=\sum_{m=1}^{n}\bm{\mu}_m$, where $n$ is the total number of ions in the primitive cell.

Based on the 2,564 materials in our database with a maximum phonon AM exceeding $0.5\hbar$, we further select those with no more than 10 atoms in the primitive cell to compute the phonon MM (workflow shown in the right panel of Fig.~\ref{Application}(a)). We identify 24 materials that host phonon modes with giant phonon MM exceeding $0.5\,\mu_N$, as listed in Table~\ref{tabPMM}.
Materials containing hydrogen tend to exhibit relatively large phonon MM because hydrogen has the smallest mass, which leads to a larger gyromagnetic ratio.
Importantly, we find that some of these phonon modes coexist with EMPs, including C-1 WP and L-NS.
This coexistence of topology and chirality may provide a promising route to enhance the phonon MM.

\section{CONCLUSION}
The successful introduction of topology and chirality as two degrees of freedom in terahertz-level phonons has provided a control handle for phononic manipulation.
In this work, we first establish a complete symmetry-based classification that shows how atomic positions contribute to the number of times (co-)irreps appear and the phonon AM of each phonon mode.
Our symmetry-based methods for analyzing phonon AM can be extended to other physical properties, such as spin texture and Berry curvature.
Moreover, we apply our symmetry results to establish the database, mapping EMPs and phonon AM to specific phononic band crossings in real materials.
Our database serves as an effective platform for the aforementioned studies and facilitates convenient exploration of the coexistence of topological and chiral phonons.
Recently, new experimental approaches have been developed to detect phonons. We believe that our work will attract numerous subsequent experimental and theoretical studies, and the large number of topological and chiral phonon materials we proposed will catalyze the discovery of new candidates.

\section{ACKNOWLEDGMENTS}
This paper was supported by the National Natural Science Foundation of China (NSFC) under Grants No. 12188101 and No. 12322404; the National Key R\&D Program of China (Grant No. 2025YFA1411301 and Grant No. 2022YFA1403601); Innovation Program for Quantum Science and Technology (Grant No. 2021ZD0301902), Quantum Science and Technology-National Science and Technology Major Project (Grant No. 2024ZD0300101); Fundamental and Interdisciplinary Disciplines Breakthrough Plan of the Ministry of Education of China (JYB2025XDXM411); the Fundamental Research Funds for the Central Universities (KG202501); and the Natural Science Foundation of Jiangsu Province (Grants No.BK20233001 and No. BK20243011). X.W. also acknowledges support from the Tencent Foundation through the XPLORER PRIZE and the New Cornerstone Science Foundation.


\begin{thebibliography}{87}%
\makeatletter
\providecommand \@ifxundefined [1]{%
 \@ifx{#1\undefined}
}%
\providecommand \@ifnum [1]{%
 \ifnum #1\expandafter \@firstoftwo
 \else \expandafter \@secondoftwo
 \fi
}%
\providecommand \@ifx [1]{%
 \ifx #1\expandafter \@firstoftwo
 \else \expandafter \@secondoftwo
 \fi
}%
\providecommand \natexlab [1]{#1}%
\providecommand \enquote  [1]{``#1''}%
\providecommand \bibnamefont  [1]{#1}%
\providecommand \bibfnamefont [1]{#1}%
\providecommand \citenamefont [1]{#1}%
\providecommand \href@noop [0]{\@secondoftwo}%
\providecommand \href [0]{\begingroup \@sanitize@url \@href}%
\providecommand \@href[1]{\@@startlink{#1}\@@href}%
\providecommand \@@href[1]{\endgroup#1\@@endlink}%
\providecommand \@sanitize@url [0]{\catcode `\\12\catcode `\$12\catcode
  `\&12\catcode `\#12\catcode `\^12\catcode `\_12\catcode `\%12\relax}%
\providecommand \@@startlink[1]{}%
\providecommand \@@endlink[0]{}%
\providecommand \url  [0]{\begingroup\@sanitize@url \@url }%
\providecommand \@url [1]{\endgroup\@href {#1}{\urlprefix }}%
\providecommand \urlprefix  [0]{URL }%
\providecommand \Eprint [0]{\href }%
\providecommand \doibase [0]{https://doi.org/}%
\providecommand \selectlanguage [0]{\@gobble}%
\providecommand \bibinfo  [0]{\@secondoftwo}%
\providecommand \bibfield  [0]{\@secondoftwo}%
\providecommand \translation [1]{[#1]}%
\providecommand \BibitemOpen [0]{}%
\providecommand \bibitemStop [0]{}%
\providecommand \bibitemNoStop [0]{.\EOS\space}%
\providecommand \EOS [0]{\spacefactor3000\relax}%
\providecommand \BibitemShut  [1]{\csname bibitem#1\endcsname}%
\let\auto@bib@innerbib\@empty
\bibitem [{\citenamefont {J.R.}(2000)}]{topo-book}%
  \BibitemOpen
  \bibfield  {author} {\bibinfo {author} {\bibfnamefont {M.}~\bibnamefont
  {J.R.}},\ }\href@noop {} {\emph {\bibinfo {title} {{Topology}}}}\ (\bibinfo
  {publisher} {Prentice Hall},\ \bibinfo {year} {2000})\BibitemShut {NoStop}%
\bibitem [{\citenamefont {Kelvin}(1894)}]{chiral-book}%
  \BibitemOpen
  \bibfield  {author} {\bibinfo {author} {\bibfnamefont {W.~T.~B.}\
  \bibnamefont {Kelvin}},\ }\href@noop {} {\emph {\bibinfo {title} {{The
  molecular tactics of a crystal}}}}\ (\bibinfo  {publisher} {Clarendon
  Press},\ \bibinfo {year} {1894})\BibitemShut {NoStop}%
\bibitem [{\citenamefont {Bradlyn}\ \emph {et~al.}(2017)\citenamefont
  {Bradlyn}, \citenamefont {Elcoro}, \citenamefont {Cano}, \citenamefont
  {Vergniory}, \citenamefont {Wang}, \citenamefont {Felser}, \citenamefont
  {Aroyo},\ and\ \citenamefont {Bernevig}}]{symmetry1}%
  \BibitemOpen
  \bibfield  {author} {\bibinfo {author} {\bibfnamefont {B.}~\bibnamefont
  {Bradlyn}}, \bibinfo {author} {\bibfnamefont {L.}~\bibnamefont {Elcoro}},
  \bibinfo {author} {\bibfnamefont {J.}~\bibnamefont {Cano}}, \bibinfo {author}
  {\bibfnamefont {M.~G.}\ \bibnamefont {Vergniory}}, \bibinfo {author}
  {\bibfnamefont {Z.}~\bibnamefont {Wang}}, \bibinfo {author} {\bibfnamefont
  {C.}~\bibnamefont {Felser}}, \bibinfo {author} {\bibfnamefont {M.~I.}\
  \bibnamefont {Aroyo}},\ and\ \bibinfo {author} {\bibfnamefont {B.~A.}\
  \bibnamefont {Bernevig}},\ }\bibfield  {title} {\bibinfo {title}
  {{Topological quantum chemistry}},\ }\href
  {https://doi.org/10.1038/nature23268} {\bibfield  {journal} {\bibinfo
  {journal} {Nature (London)}\ }\textbf {\bibinfo {volume} {547}},\ \bibinfo
  {pages} {298} (\bibinfo {year} {2017})}\BibitemShut {NoStop}%
\bibitem [{\citenamefont {Elcoro}\ \emph {et~al.}(2021)\citenamefont {Elcoro},
  \citenamefont {Wieder}, \citenamefont {Song}, \citenamefont {Xu},
  \citenamefont {Bradlyn},\ and\ \citenamefont {Bernevig}}]{symmetry2}%
  \BibitemOpen
  \bibfield  {author} {\bibinfo {author} {\bibfnamefont {L.}~\bibnamefont
  {Elcoro}}, \bibinfo {author} {\bibfnamefont {B.~J.}\ \bibnamefont {Wieder}},
  \bibinfo {author} {\bibfnamefont {Z.}~\bibnamefont {Song}}, \bibinfo {author}
  {\bibfnamefont {Y.}~\bibnamefont {Xu}}, \bibinfo {author} {\bibfnamefont
  {B.}~\bibnamefont {Bradlyn}},\ and\ \bibinfo {author} {\bibfnamefont {B.~A.}\
  \bibnamefont {Bernevig}},\ }\bibfield  {title} {\bibinfo {title} {Magnetic
  topological quantum chemistry},\ }\href
  {https://doi.org/10.1038/s41467-021-26241-8} {\bibfield  {journal} {\bibinfo
  {journal} {Nature Communications}\ }\textbf {\bibinfo {volume} {12}},\
  \bibinfo {pages} {5965} (\bibinfo {year} {2021})}\BibitemShut {NoStop}%
\bibitem [{\citenamefont {Fu}\ and\ \citenamefont {Kane}(2006)}]{symmetry3}%
  \BibitemOpen
  \bibfield  {author} {\bibinfo {author} {\bibfnamefont {L.}~\bibnamefont
  {Fu}}\ and\ \bibinfo {author} {\bibfnamefont {C.~L.}\ \bibnamefont {Kane}},\
  }\bibfield  {title} {\bibinfo {title} {{Time reversal polarization and a
  ${Z}_{2}$ adiabatic spin pump}},\ }\href
  {https://doi.org/10.1103/PhysRevB.74.195312} {\bibfield  {journal} {\bibinfo
  {journal} {Phys. Rev. B}\ }\textbf {\bibinfo {volume} {74}},\ \bibinfo
  {pages} {195312} (\bibinfo {year} {2006})}\BibitemShut {NoStop}%
\bibitem [{\citenamefont {Po}\ \emph {et~al.}(2017)\citenamefont {Po},
  \citenamefont {Vishwanath},\ and\ \citenamefont {Watanabe}}]{symmetry4}%
  \BibitemOpen
  \bibfield  {author} {\bibinfo {author} {\bibfnamefont {H.~C.}\ \bibnamefont
  {Po}}, \bibinfo {author} {\bibfnamefont {A.}~\bibnamefont {Vishwanath}},\
  and\ \bibinfo {author} {\bibfnamefont {H.}~\bibnamefont {Watanabe}},\
  }\bibfield  {title} {\bibinfo {title} {Symmetry-based indicators of band
  topology in the 230 space groups},\ }\href
  {https://doi.org/10.1038/s41467-017-00133-2} {\bibfield  {journal} {\bibinfo
  {journal} {Nat. Commun.}\ }\textbf {\bibinfo {volume} {8}},\ \bibinfo {pages}
  {50} (\bibinfo {year} {2017})}\BibitemShut {NoStop}%
\bibitem [{\citenamefont {Song}\ \emph
  {et~al.}(2018{\natexlab{a}})\citenamefont {Song}, \citenamefont {Zhang},
  \citenamefont {Fang},\ and\ \citenamefont {Fang}}]{symmetry5}%
  \BibitemOpen
  \bibfield  {author} {\bibinfo {author} {\bibfnamefont {Z.}~\bibnamefont
  {Song}}, \bibinfo {author} {\bibfnamefont {T.}~\bibnamefont {Zhang}},
  \bibinfo {author} {\bibfnamefont {Z.}~\bibnamefont {Fang}},\ and\ \bibinfo
  {author} {\bibfnamefont {C.}~\bibnamefont {Fang}},\ }\bibfield  {title}
  {\bibinfo {title} {Quantitative mappings between symmetry and topology in
  solids},\ }\href {https://doi.org/10.1038/s41467-018-06010-w} {\bibfield
  {journal} {\bibinfo  {journal} {Nature Communications}\ }\textbf {\bibinfo
  {volume} {9}},\ \bibinfo {pages} {3530} (\bibinfo {year}
  {2018}{\natexlab{a}})}\BibitemShut {NoStop}%
\bibitem [{\citenamefont {{Haruki Watanabe and Hoi Chun Po and Ashvin
  Vishwanath}}(2018)}]{symmetry6}%
  \BibitemOpen
  \bibfield  {author} {\bibinfo {author} {\bibnamefont {{Haruki Watanabe and
  Hoi Chun Po and Ashvin Vishwanath}}},\ }\bibfield  {title} {\bibinfo {title}
  {{Structure and topology of band structures in the 1651 magnetic space
  groups}},\ }\href {https://doi.org/10.1126/sciadv.aat8685} {\bibfield
  {journal} {\bibinfo  {journal} {Science Advances}\ }\textbf {\bibinfo
  {volume} {4}},\ \bibinfo {pages} {eaat8685} (\bibinfo {year}
  {2018})}\BibitemShut {NoStop}%
\bibitem [{\citenamefont {Slager}\ \emph {et~al.}(2013)\citenamefont {Slager},
  \citenamefont {Mesaros}, \citenamefont {Juri\v{c}i\'{c}},\ and\ \citenamefont
  {Zaanen}}]{symmetry7}%
  \BibitemOpen
  \bibfield  {author} {\bibinfo {author} {\bibfnamefont {R.-J.}\ \bibnamefont
  {Slager}}, \bibinfo {author} {\bibfnamefont {A.}~\bibnamefont {Mesaros}},
  \bibinfo {author} {\bibfnamefont {V.}~\bibnamefont {Juri\v{c}i\'{c}}},\ and\
  \bibinfo {author} {\bibfnamefont {J.}~\bibnamefont {Zaanen}},\ }\bibfield
  {title} {\bibinfo {title} {The space group classification of topological
  band-insulators},\ }\href {https://doi.org/10.1038/nphys2513} {\bibfield
  {journal} {\bibinfo  {journal} {Nature Physics}\ }\textbf {\bibinfo {volume}
  {9}},\ \bibinfo {pages} {98} (\bibinfo {year} {2013})}\BibitemShut {NoStop}%
\bibitem [{\citenamefont {Kruthoff}\ \emph {et~al.}(2017)\citenamefont
  {Kruthoff}, \citenamefont {de~Boer}, \citenamefont {van Wezel}, \citenamefont
  {Kane},\ and\ \citenamefont {Slager}}]{symmetry8}%
  \BibitemOpen
  \bibfield  {author} {\bibinfo {author} {\bibfnamefont {J.}~\bibnamefont
  {Kruthoff}}, \bibinfo {author} {\bibfnamefont {J.}~\bibnamefont {de~Boer}},
  \bibinfo {author} {\bibfnamefont {J.}~\bibnamefont {van Wezel}}, \bibinfo
  {author} {\bibfnamefont {C.~L.}\ \bibnamefont {Kane}},\ and\ \bibinfo
  {author} {\bibfnamefont {R.-J.}\ \bibnamefont {Slager}},\ }\bibfield  {title}
  {\bibinfo {title} {{Topological Classification of Crystalline Insulators
  through Band Structure Combinatorics}},\ }\href
  {https://doi.org/10.1103/PhysRevX.7.041069} {\bibfield  {journal} {\bibinfo
  {journal} {Phys. Rev. X}\ }\textbf {\bibinfo {volume} {7}},\ \bibinfo {pages}
  {041069} (\bibinfo {year} {2017})}\BibitemShut {NoStop}%
\bibitem [{\citenamefont {Zhang}\ \emph {et~al.}(2019)\citenamefont {Zhang},
  \citenamefont {Jiang}, \citenamefont {Song}, \citenamefont {Huang},
  \citenamefont {He}, \citenamefont {Fang}, \citenamefont {Weng},\ and\
  \citenamefont {Fang}}]{catalogue1}%
  \BibitemOpen
  \bibfield  {author} {\bibinfo {author} {\bibfnamefont {T.}~\bibnamefont
  {Zhang}}, \bibinfo {author} {\bibfnamefont {Y.}~\bibnamefont {Jiang}},
  \bibinfo {author} {\bibfnamefont {Z.}~\bibnamefont {Song}}, \bibinfo {author}
  {\bibfnamefont {H.}~\bibnamefont {Huang}}, \bibinfo {author} {\bibfnamefont
  {Y.}~\bibnamefont {He}}, \bibinfo {author} {\bibfnamefont {Z.}~\bibnamefont
  {Fang}}, \bibinfo {author} {\bibfnamefont {H.}~\bibnamefont {Weng}},\ and\
  \bibinfo {author} {\bibfnamefont {C.}~\bibnamefont {Fang}},\ }\bibfield
  {title} {\bibinfo {title} {{Catalogue of topological electronic materials}},\
  }\href {https://doi.org/10.1038/s41586-019-0944-6} {\bibfield  {journal}
  {\bibinfo  {journal} {Nature (London)}\ }\textbf {\bibinfo {volume} {566}},\
  \bibinfo {pages} {475} (\bibinfo {year} {2019})}\BibitemShut {NoStop}%
\bibitem [{\citenamefont {Vergniory}\ \emph {et~al.}(2019)\citenamefont
  {Vergniory}, \citenamefont {Elcoro}, \citenamefont {Felser}, \citenamefont
  {Regnault}, \citenamefont {Bernevig},\ and\ \citenamefont
  {Wang}}]{catalogue2}%
  \BibitemOpen
  \bibfield  {author} {\bibinfo {author} {\bibfnamefont {M.~G.}\ \bibnamefont
  {Vergniory}}, \bibinfo {author} {\bibfnamefont {L.}~\bibnamefont {Elcoro}},
  \bibinfo {author} {\bibfnamefont {C.}~\bibnamefont {Felser}}, \bibinfo
  {author} {\bibfnamefont {N.}~\bibnamefont {Regnault}}, \bibinfo {author}
  {\bibfnamefont {B.~A.}\ \bibnamefont {Bernevig}},\ and\ \bibinfo {author}
  {\bibfnamefont {Z.}~\bibnamefont {Wang}},\ }\bibfield  {title} {\bibinfo
  {title} {{A complete catalogue of high-quality topological materials}},\
  }\href {https://doi.org/10.1038/s41586-019-0954-4} {\bibfield  {journal}
  {\bibinfo  {journal} {Nature (London)}\ }\textbf {\bibinfo {volume} {566}},\
  \bibinfo {pages} {480} (\bibinfo {year} {2019})}\BibitemShut {NoStop}%
\bibitem [{\citenamefont {Tang}\ \emph
  {et~al.}(2019{\natexlab{a}})\citenamefont {Tang}, \citenamefont {Po},
  \citenamefont {Vishwanath},\ and\ \citenamefont {Wan}}]{catalogue3}%
  \BibitemOpen
  \bibfield  {author} {\bibinfo {author} {\bibfnamefont {F.}~\bibnamefont
  {Tang}}, \bibinfo {author} {\bibfnamefont {H.~C.}\ \bibnamefont {Po}},
  \bibinfo {author} {\bibfnamefont {A.}~\bibnamefont {Vishwanath}},\ and\
  \bibinfo {author} {\bibfnamefont {X.}~\bibnamefont {Wan}},\ }\bibfield
  {title} {\bibinfo {title} {{Comprehensive search for topological materials
  using symmetry indicators}},\ }\href
  {https://doi.org/10.1038/s41586-019-0937-5} {\bibfield  {journal} {\bibinfo
  {journal} {Nature (London)}\ }\textbf {\bibinfo {volume} {566}},\ \bibinfo
  {pages} {486} (\bibinfo {year} {2019}{\natexlab{a}})}\BibitemShut {NoStop}%
\bibitem [{\citenamefont {Vergniory}\ \emph
  {et~al.}(2022{\natexlab{a}})\citenamefont {Vergniory}, \citenamefont
  {Wieder}, \citenamefont {Elcoro}, \citenamefont {Parkin}, \citenamefont
  {Felser}, \citenamefont {Bernevig},\ and\ \citenamefont
  {Regnault}}]{catalogue4}%
  \BibitemOpen
  \bibfield  {author} {\bibinfo {author} {\bibfnamefont {M.~G.}\ \bibnamefont
  {Vergniory}}, \bibinfo {author} {\bibfnamefont {B.~J.}\ \bibnamefont
  {Wieder}}, \bibinfo {author} {\bibfnamefont {L.}~\bibnamefont {Elcoro}},
  \bibinfo {author} {\bibfnamefont {S.~S.~P.}\ \bibnamefont {Parkin}}, \bibinfo
  {author} {\bibfnamefont {C.}~\bibnamefont {Felser}}, \bibinfo {author}
  {\bibfnamefont {B.~A.}\ \bibnamefont {Bernevig}},\ and\ \bibinfo {author}
  {\bibfnamefont {N.}~\bibnamefont {Regnault}},\ }\bibfield  {title} {\bibinfo
  {title} {{All topological bands of all nonmagnetic stoichiometric
  materials}},\ }\href {https://doi.org/10.1126/science.abg9094} {\bibfield
  {journal} {\bibinfo  {journal} {Science}\ }\textbf {\bibinfo {volume}
  {376}},\ \bibinfo {pages} {eabg9094} (\bibinfo {year}
  {2022}{\natexlab{a}})}\BibitemShut {NoStop}%
\bibitem [{\citenamefont {Song}\ \emph
  {et~al.}(2018{\natexlab{b}})\citenamefont {Song}, \citenamefont {Zhang},\
  and\ \citenamefont {Fang}}]{catalogue5}%
  \BibitemOpen
  \bibfield  {author} {\bibinfo {author} {\bibfnamefont {Z.}~\bibnamefont
  {Song}}, \bibinfo {author} {\bibfnamefont {T.}~\bibnamefont {Zhang}},\ and\
  \bibinfo {author} {\bibfnamefont {C.}~\bibnamefont {Fang}},\ }\bibfield
  {title} {\bibinfo {title} {{Diagnosis for Nonmagnetic Topological Semimetals
  in the Absence of Spin-Orbital Coupling}},\ }\href
  {https://doi.org/10.1103/PhysRevX.8.031069} {\bibfield  {journal} {\bibinfo
  {journal} {Phys. Rev. X}\ }\textbf {\bibinfo {volume} {8}},\ \bibinfo {pages}
  {031069} (\bibinfo {year} {2018}{\natexlab{b}})}\BibitemShut {NoStop}%
\bibitem [{\citenamefont {Tang}\ \emph
  {et~al.}(2019{\natexlab{b}})\citenamefont {Tang}, \citenamefont {Po},
  \citenamefont {Vishwanath},\ and\ \citenamefont {Wan}}]{catalogue6}%
  \BibitemOpen
  \bibfield  {author} {\bibinfo {author} {\bibfnamefont {F.}~\bibnamefont
  {Tang}}, \bibinfo {author} {\bibfnamefont {H.~C.}\ \bibnamefont {Po}},
  \bibinfo {author} {\bibfnamefont {A.}~\bibnamefont {Vishwanath}},\ and\
  \bibinfo {author} {\bibfnamefont {X.}~\bibnamefont {Wan}},\ }\bibfield
  {title} {\bibinfo {title} {Efficient topological materials discovery using
  symmetry indicators},\ }\href {https://doi.org/10.1038/s41567-019-0418-7}
  {\bibfield  {journal} {\bibinfo  {journal} {Nature Physics}\ }\textbf
  {\bibinfo {volume} {15}},\ \bibinfo {pages} {470} (\bibinfo {year}
  {2019}{\natexlab{b}})}\BibitemShut {NoStop}%
\bibitem [{\citenamefont {Tang}\ \emph
  {et~al.}(2019{\natexlab{c}})\citenamefont {Tang}, \citenamefont {Po},
  \citenamefont {Vishwanath},\ and\ \citenamefont {Wan}}]{catalogue7}%
  \BibitemOpen
  \bibfield  {author} {\bibinfo {author} {\bibfnamefont {F.}~\bibnamefont
  {Tang}}, \bibinfo {author} {\bibfnamefont {H.~C.}\ \bibnamefont {Po}},
  \bibinfo {author} {\bibfnamefont {A.}~\bibnamefont {Vishwanath}},\ and\
  \bibinfo {author} {\bibfnamefont {X.}~\bibnamefont {Wan}},\ }\bibfield
  {title} {\bibinfo {title} {{Topological materials discovery by large-order
  symmetry indicators}},\ }\href {https://doi.org/10.1126/sciadv.aau8725}
  {\bibfield  {journal} {\bibinfo  {journal} {Science Advances}\ }\textbf
  {\bibinfo {volume} {5}},\ \bibinfo {pages} {eaau8725} (\bibinfo {year}
  {2019}{\natexlab{c}})}\BibitemShut {NoStop}%
\bibitem [{\citenamefont {Wang}\ \emph {et~al.}(2019)\citenamefont {Wang},
  \citenamefont {Tang}, \citenamefont {Ji}, \citenamefont {Zhang},
  \citenamefont {Vishwanath}, \citenamefont {Po},\ and\ \citenamefont
  {Wan}}]{catalogue8}%
  \BibitemOpen
  \bibfield  {author} {\bibinfo {author} {\bibfnamefont {D.}~\bibnamefont
  {Wang}}, \bibinfo {author} {\bibfnamefont {F.}~\bibnamefont {Tang}}, \bibinfo
  {author} {\bibfnamefont {J.}~\bibnamefont {Ji}}, \bibinfo {author}
  {\bibfnamefont {W.}~\bibnamefont {Zhang}}, \bibinfo {author} {\bibfnamefont
  {A.}~\bibnamefont {Vishwanath}}, \bibinfo {author} {\bibfnamefont {H.~C.}\
  \bibnamefont {Po}},\ and\ \bibinfo {author} {\bibfnamefont {X.}~\bibnamefont
  {Wan}},\ }\bibfield  {title} {\bibinfo {title} {Two-dimensional topological
  materials discovery by symmetry-indicator method},\ }\href
  {https://doi.org/10.1103/PhysRevB.100.195108} {\bibfield  {journal} {\bibinfo
   {journal} {Phys. Rev. B}\ }\textbf {\bibinfo {volume} {100}},\ \bibinfo
  {pages} {195108} (\bibinfo {year} {2019})}\BibitemShut {NoStop}%
\bibitem [{\citenamefont {Xu}\ \emph {et~al.}(2020)\citenamefont {Xu},
  \citenamefont {Elcoro}, \citenamefont {Song}, \citenamefont {Wieder},
  \citenamefont {Vergniory}, \citenamefont {Regnault}, \citenamefont {Chen},
  \citenamefont {Felser},\ and\ \citenamefont {Bernevig}}]{catalogue9}%
  \BibitemOpen
  \bibfield  {author} {\bibinfo {author} {\bibfnamefont {Y.}~\bibnamefont
  {Xu}}, \bibinfo {author} {\bibfnamefont {L.}~\bibnamefont {Elcoro}}, \bibinfo
  {author} {\bibfnamefont {Z.-D.}\ \bibnamefont {Song}}, \bibinfo {author}
  {\bibfnamefont {B.~J.}\ \bibnamefont {Wieder}}, \bibinfo {author}
  {\bibfnamefont {M.~G.}\ \bibnamefont {Vergniory}}, \bibinfo {author}
  {\bibfnamefont {N.}~\bibnamefont {Regnault}}, \bibinfo {author}
  {\bibfnamefont {Y.}~\bibnamefont {Chen}}, \bibinfo {author} {\bibfnamefont
  {C.}~\bibnamefont {Felser}},\ and\ \bibinfo {author} {\bibfnamefont {B.~A.}\
  \bibnamefont {Bernevig}},\ }\bibfield  {title} {\bibinfo {title}
  {High-throughput calculations of magnetic topological materials},\ }\href
  {https://doi.org/10.1038/s41586-020-2837-0} {\bibfield  {journal} {\bibinfo
  {journal} {Nature}\ }\textbf {\bibinfo {volume} {586}},\ \bibinfo {pages}
  {702} (\bibinfo {year} {2020})}\BibitemShut {NoStop}%
\bibitem [{\citenamefont {Peng}\ \emph {et~al.}(2022)\citenamefont {Peng},
  \citenamefont {Jiang}, \citenamefont {Fang}, \citenamefont {Weng},\ and\
  \citenamefont {Fang}}]{catalogue10}%
  \BibitemOpen
  \bibfield  {author} {\bibinfo {author} {\bibfnamefont {B.}~\bibnamefont
  {Peng}}, \bibinfo {author} {\bibfnamefont {Y.}~\bibnamefont {Jiang}},
  \bibinfo {author} {\bibfnamefont {Z.}~\bibnamefont {Fang}}, \bibinfo {author}
  {\bibfnamefont {H.}~\bibnamefont {Weng}},\ and\ \bibinfo {author}
  {\bibfnamefont {C.}~\bibnamefont {Fang}},\ }\bibfield  {title} {\bibinfo
  {title} {{Topological classification and diagnosis in magnetically ordered
  electronic materials}},\ }\href {https://doi.org/10.1103/PhysRevB.105.235138}
  {\bibfield  {journal} {\bibinfo  {journal} {Phys. Rev. B}\ }\textbf {\bibinfo
  {volume} {105}},\ \bibinfo {pages} {235138} (\bibinfo {year}
  {2022})}\BibitemShut {NoStop}%
\bibitem [{\citenamefont {Tang}\ and\ \citenamefont
  {Wan}(2024{\natexlab{a}})}]{catalogue11}%
  \BibitemOpen
  \bibfield  {author} {\bibinfo {author} {\bibfnamefont {F.}~\bibnamefont
  {Tang}}\ and\ \bibinfo {author} {\bibfnamefont {X.}~\bibnamefont {Wan}},\
  }\bibfield  {title} {\bibinfo {title} {Topological state evolution by
  symmetry-breaking},\ }\href@noop {} {\  (\bibinfo {year}
  {2024}{\natexlab{a}})},\ \Eprint {https://arxiv.org/abs/2302.13622}
  {arXiv:2302.13622} \BibitemShut {NoStop}%
\bibitem [{\citenamefont {Robredo}\ \emph {et~al.}(2025)\citenamefont
  {Robredo}, \citenamefont {Xu}, \citenamefont {Jiang}, \citenamefont {Felser},
  \citenamefont {Bernevig}, \citenamefont {Elcoro}, \citenamefont {Regnault},\
  and\ \citenamefont {Vergniory}}]{catalogue12}%
  \BibitemOpen
  \bibfield  {author} {\bibinfo {author} {\bibfnamefont {I.}~\bibnamefont
  {Robredo}}, \bibinfo {author} {\bibfnamefont {Y.}~\bibnamefont {Xu}},
  \bibinfo {author} {\bibfnamefont {Y.}~\bibnamefont {Jiang}}, \bibinfo
  {author} {\bibfnamefont {C.}~\bibnamefont {Felser}}, \bibinfo {author}
  {\bibfnamefont {B.~A.}\ \bibnamefont {Bernevig}}, \bibinfo {author}
  {\bibfnamefont {L.}~\bibnamefont {Elcoro}}, \bibinfo {author} {\bibfnamefont
  {N.}~\bibnamefont {Regnault}},\ and\ \bibinfo {author} {\bibfnamefont
  {M.~G.}\ \bibnamefont {Vergniory}},\ }\bibfield  {title} {\bibinfo {title}
  {New magnetic topological materials from high-throughput search},\ }\href
  {https://doi.org/10.1126/sciadv.adv8780} {\bibfield  {journal} {\bibinfo
  {journal} {Science Advances}\ }\textbf {\bibinfo {volume} {11}},\ \bibinfo
  {pages} {eadv8780} (\bibinfo {year} {2025})}\BibitemShut {NoStop}%
\bibitem [{\citenamefont {Vergniory}\ \emph
  {et~al.}(2022{\natexlab{b}})\citenamefont {Vergniory}, \citenamefont
  {Wieder}, \citenamefont {Elcoro}, \citenamefont {Parkin}, \citenamefont
  {Felser}, \citenamefont {Bernevig},\ and\ \citenamefont
  {Regnault}}]{catalogue13}%
  \BibitemOpen
  \bibfield  {author} {\bibinfo {author} {\bibfnamefont {M.~G.}\ \bibnamefont
  {Vergniory}}, \bibinfo {author} {\bibfnamefont {B.~J.}\ \bibnamefont
  {Wieder}}, \bibinfo {author} {\bibfnamefont {L.}~\bibnamefont {Elcoro}},
  \bibinfo {author} {\bibfnamefont {S.~S.~P.}\ \bibnamefont {Parkin}}, \bibinfo
  {author} {\bibfnamefont {C.}~\bibnamefont {Felser}}, \bibinfo {author}
  {\bibfnamefont {B.~A.}\ \bibnamefont {Bernevig}},\ and\ \bibinfo {author}
  {\bibfnamefont {N.}~\bibnamefont {Regnault}},\ }\bibfield  {title} {\bibinfo
  {title} {{All topological bands of all nonmagnetic stoichiometric
  materials}},\ }\href {https://doi.org/10.1126/science.abg9094} {\bibfield
  {journal} {\bibinfo  {journal} {Science}\ }\textbf {\bibinfo {volume}
  {376}},\ \bibinfo {pages} {eabg9094} (\bibinfo {year}
  {2022}{\natexlab{b}})}\BibitemShut {NoStop}%
\bibitem [{\citenamefont {Petretto}\ \emph {et~al.}(2018)\citenamefont
  {Petretto}, \citenamefont {Dwaraknath}, \citenamefont {Miranda},
  \citenamefont {Winston}, \citenamefont {Giantomassi}, \citenamefont
  {v.~Setten}, \citenamefont {Gonze}, \citenamefont {Persson}, \citenamefont
  {Hautier},\ and\ \citenamefont {Rignanese}}]{HPCphonon1}%
  \BibitemOpen
  \bibfield  {author} {\bibinfo {author} {\bibfnamefont {G.}~\bibnamefont
  {Petretto}}, \bibinfo {author} {\bibfnamefont {S.}~\bibnamefont
  {Dwaraknath}}, \bibinfo {author} {\bibfnamefont {H.~P.}\ \bibnamefont
  {Miranda}}, \bibinfo {author} {\bibfnamefont {D.}~\bibnamefont {Winston}},
  \bibinfo {author} {\bibfnamefont {M.}~\bibnamefont {Giantomassi}}, \bibinfo
  {author} {\bibfnamefont {M.~J.}\ \bibnamefont {v.~Setten}}, \bibinfo {author}
  {\bibfnamefont {X.}~\bibnamefont {Gonze}}, \bibinfo {author} {\bibfnamefont
  {K.~A.}\ \bibnamefont {Persson}}, \bibinfo {author} {\bibfnamefont
  {G.}~\bibnamefont {Hautier}},\ and\ \bibinfo {author} {\bibfnamefont {G.-M.}\
  \bibnamefont {Rignanese}},\ }\bibfield  {title} {\bibinfo {title}
  {{High-throughput density-functional perturbation theory phonons for
  inorganic materials}},\ }\href {https://doi.org/10.1038/sdata.2018.65}
  {\bibfield  {journal} {\bibinfo  {journal} {Sci. Data}\ }\textbf {\bibinfo
  {volume} {5}},\ \bibinfo {pages} {1} (\bibinfo {year} {2018})}\BibitemShut
  {NoStop}%
\bibitem [{\citenamefont {Li}\ \emph {et~al.}(2021)\citenamefont {Li},
  \citenamefont {Liu}, \citenamefont {Baronett}, \citenamefont {Liu},
  \citenamefont {Wang}, \citenamefont {Li}, \citenamefont {Chen}, \citenamefont
  {Li}, \citenamefont {Zhu},\ and\ \citenamefont {Chen}}]{HPCphonon2}%
  \BibitemOpen
  \bibfield  {author} {\bibinfo {author} {\bibfnamefont {J.}~\bibnamefont
  {Li}}, \bibinfo {author} {\bibfnamefont {J.}~\bibnamefont {Liu}}, \bibinfo
  {author} {\bibfnamefont {S.~A.}\ \bibnamefont {Baronett}}, \bibinfo {author}
  {\bibfnamefont {M.}~\bibnamefont {Liu}}, \bibinfo {author} {\bibfnamefont
  {L.}~\bibnamefont {Wang}}, \bibinfo {author} {\bibfnamefont {R.}~\bibnamefont
  {Li}}, \bibinfo {author} {\bibfnamefont {Y.}~\bibnamefont {Chen}}, \bibinfo
  {author} {\bibfnamefont {D.}~\bibnamefont {Li}}, \bibinfo {author}
  {\bibfnamefont {Q.}~\bibnamefont {Zhu}},\ and\ \bibinfo {author}
  {\bibfnamefont {X.-Q.}\ \bibnamefont {Chen}},\ }\bibfield  {title} {\bibinfo
  {title} {{Computation and data driven discovery of topological phononic
  materials}},\ }\href {https://doi.org/10.1038/s41467-021-21293-2} {\bibfield
  {journal} {\bibinfo  {journal} {Nat. Commun.}\ }\textbf {\bibinfo {volume}
  {12}},\ \bibinfo {pages} {1204} (\bibinfo {year} {2021})}\BibitemShut
  {NoStop}%
\bibitem [{\citenamefont {Xu}\ \emph {et~al.}(2024)\citenamefont {Xu},
  \citenamefont {Vergniory}, \citenamefont {Ma}, \citenamefont {Ma{\~n}es},
  \citenamefont {Song}, \citenamefont {Bernevig}, \citenamefont {Regnault},\
  and\ \citenamefont {Elcoro}}]{HPCphonon3}%
  \BibitemOpen
  \bibfield  {author} {\bibinfo {author} {\bibfnamefont {Y.}~\bibnamefont
  {Xu}}, \bibinfo {author} {\bibfnamefont {M.~G.}\ \bibnamefont {Vergniory}},
  \bibinfo {author} {\bibfnamefont {D.-S.}\ \bibnamefont {Ma}}, \bibinfo
  {author} {\bibfnamefont {J.~L.}\ \bibnamefont {Ma{\~n}es}}, \bibinfo {author}
  {\bibfnamefont {Z.-D.}\ \bibnamefont {Song}}, \bibinfo {author}
  {\bibfnamefont {B.~A.}\ \bibnamefont {Bernevig}}, \bibinfo {author}
  {\bibfnamefont {N.}~\bibnamefont {Regnault}},\ and\ \bibinfo {author}
  {\bibfnamefont {L.}~\bibnamefont {Elcoro}},\ }\bibfield  {title} {\bibinfo
  {title} {{Catalog of topological phonon materials}},\ }\href
  {https://doi.org/10.1126/science.adf8458} {\bibfield  {journal} {\bibinfo
  {journal} {Science}\ }\textbf {\bibinfo {volume} {384}},\ \bibinfo {pages}
  {eadf8458} (\bibinfo {year} {2024})}\BibitemShut {NoStop}%
\bibitem [{\citenamefont {Tang}\ and\ \citenamefont {Wan}(2021)}]{tangkp1}%
  \BibitemOpen
  \bibfield  {author} {\bibinfo {author} {\bibfnamefont {F.}~\bibnamefont
  {Tang}}\ and\ \bibinfo {author} {\bibfnamefont {X.}~\bibnamefont {Wan}},\
  }\bibfield  {title} {\bibinfo {title} {{Exhaustive construction of effective
  models in 1651 magnetic space groups}},\ }\href
  {https://doi.org/10.1103/PhysRevB.104.085137} {\bibfield  {journal} {\bibinfo
   {journal} {Phys. Rev. B}\ }\textbf {\bibinfo {volume} {104}},\ \bibinfo
  {pages} {085137} (\bibinfo {year} {2021})}\BibitemShut {NoStop}%
\bibitem [{\citenamefont {Tang}\ and\ \citenamefont {Wan}(2022)}]{tangkp2}%
  \BibitemOpen
  \bibfield  {author} {\bibinfo {author} {\bibfnamefont {F.}~\bibnamefont
  {Tang}}\ and\ \bibinfo {author} {\bibfnamefont {X.}~\bibnamefont {Wan}},\
  }\bibfield  {title} {\bibinfo {title} {{Complete classification of band nodal
  structures and massless excitations}},\ }\href
  {https://doi.org/10.1103/PhysRevB.105.155156} {\bibfield  {journal} {\bibinfo
   {journal} {Phys. Rev. B}\ }\textbf {\bibinfo {volume} {105}},\ \bibinfo
  {pages} {155156} (\bibinfo {year} {2022})}\BibitemShut {NoStop}%
\bibitem [{\citenamefont {Yu}\ \emph {et~al.}(2022)\citenamefont {Yu},
  \citenamefont {Zhang}, \citenamefont {Liu}, \citenamefont {Wu}, \citenamefont
  {Li}, \citenamefont {Zhang}, \citenamefont {Yang},\ and\ \citenamefont
  {Yao}}]{SciBul-yao}%
  \BibitemOpen
  \bibfield  {author} {\bibinfo {author} {\bibfnamefont {Z.-M.}\ \bibnamefont
  {Yu}}, \bibinfo {author} {\bibfnamefont {Z.}~\bibnamefont {Zhang}}, \bibinfo
  {author} {\bibfnamefont {G.-B.}\ \bibnamefont {Liu}}, \bibinfo {author}
  {\bibfnamefont {W.}~\bibnamefont {Wu}}, \bibinfo {author} {\bibfnamefont
  {X.-P.}\ \bibnamefont {Li}}, \bibinfo {author} {\bibfnamefont {R.-W.}\
  \bibnamefont {Zhang}}, \bibinfo {author} {\bibfnamefont {S.~A.}\ \bibnamefont
  {Yang}},\ and\ \bibinfo {author} {\bibfnamefont {Y.}~\bibnamefont {Yao}},\
  }\bibfield  {title} {\bibinfo {title} {{Encyclopedia of emergent particles in
  three-dimensional crystals}},\ }\href
  {https://doi.org/https://doi.org/10.1016/j.scib.2021.10.023} {\bibfield
  {journal} {\bibinfo  {journal} {Sci. Bull.}\ }\textbf {\bibinfo {volume}
  {67}},\ \bibinfo {pages} {375} (\bibinfo {year} {2022})}\BibitemShut
  {NoStop}%
\bibitem [{\citenamefont {Shekhar}(2018)}]{TopoChiral1}%
  \BibitemOpen
  \bibfield  {author} {\bibinfo {author} {\bibfnamefont {C.}~\bibnamefont
  {Shekhar}},\ }\bibfield  {title} {\bibinfo {title} {Chirality meets
  topology},\ }\href {https://doi.org/10.1038/s41563-018-0210-6} {\bibfield
  {journal} {\bibinfo  {journal} {Nature Materials}\ }\textbf {\bibinfo
  {volume} {17}},\ \bibinfo {pages} {953} (\bibinfo {year} {2018})}\BibitemShut
  {NoStop}%
\bibitem [{\citenamefont {Felser}\ and\ \citenamefont {Gooth}()}]{TopoChiral2}%
  \BibitemOpen
  \bibfield  {author} {\bibinfo {author} {\bibfnamefont {C.}~\bibnamefont
  {Felser}}\ and\ \bibinfo {author} {\bibfnamefont {J.}~\bibnamefont {Gooth}},\
  }\bibinfo {title} {Topology and chirality},\ in\ \href
  {https://doi.org/10.1142/9789811265068_0010} {\emph {\bibinfo {booktitle}
  {Chiral Matter}}},\ p.\ \bibinfo {pages} {115}\BibitemShut {NoStop}%
\bibitem [{\citenamefont {Wu}\ \emph {et~al.}(2025)\citenamefont {Wu},
  \citenamefont {Wang},\ and\ \citenamefont {Felser}}]{TopoChiral3}%
  \BibitemOpen
  \bibfield  {author} {\bibinfo {author} {\bibfnamefont {X.}~\bibnamefont
  {Wu}}, \bibinfo {author} {\bibfnamefont {X.}~\bibnamefont {Wang}},\ and\
  \bibinfo {author} {\bibfnamefont {C.}~\bibnamefont {Felser}},\ }\bibfield
  {title} {\bibinfo {title} {{Chirality meets topology: building quantum
  bridges to catalysis}},\ }\href {https://doi.org/10.1007/s40766-025-00068-1}
  {\bibfield  {journal} {\bibinfo  {journal} {RIVISTA DEL NUOVO CIMENTO}\
  }\textbf {\bibinfo {volume} {48}},\ \bibinfo {pages} {241} (\bibinfo {year}
  {2025})}\BibitemShut {NoStop}%
\bibitem [{\citenamefont {Klitzing}\ \emph {et~al.}(1980)\citenamefont
  {Klitzing}, \citenamefont {Dorda},\ and\ \citenamefont
  {Pepper}}]{Hallsystem1}%
  \BibitemOpen
  \bibfield  {author} {\bibinfo {author} {\bibfnamefont {K.~v.}\ \bibnamefont
  {Klitzing}}, \bibinfo {author} {\bibfnamefont {G.}~\bibnamefont {Dorda}},\
  and\ \bibinfo {author} {\bibfnamefont {M.}~\bibnamefont {Pepper}},\
  }\bibfield  {title} {\bibinfo {title} {{New Method for High-Accuracy
  Determination of the Fine-Structure Constant Based on Quantized Hall
  Resistance}},\ }\href {https://doi.org/10.1103/PhysRevLett.45.494} {\bibfield
   {journal} {\bibinfo  {journal} {Phys. Rev. Lett.}\ }\textbf {\bibinfo
  {volume} {45}},\ \bibinfo {pages} {494} (\bibinfo {year} {1980})}\BibitemShut
  {NoStop}%
\bibitem [{\citenamefont {von Klitzing}(1986)}]{Hallsystem2}%
  \BibitemOpen
  \bibfield  {author} {\bibinfo {author} {\bibfnamefont {K.}~\bibnamefont {von
  Klitzing}},\ }\bibfield  {title} {\bibinfo {title} {{The quantized Hall
  effect}},\ }\href {https://doi.org/10.1103/RevModPhys.58.519} {\bibfield
  {journal} {\bibinfo  {journal} {Rev. Mod. Phys.}\ }\textbf {\bibinfo {volume}
  {58}},\ \bibinfo {pages} {519} (\bibinfo {year} {1986})}\BibitemShut
  {NoStop}%
\bibitem [{\citenamefont {Wan}\ \emph {et~al.}(2011)\citenamefont {Wan},
  \citenamefont {Turner}, \citenamefont {Vishwanath},\ and\ \citenamefont
  {Savrasov}}]{Weyl-Wan}%
  \BibitemOpen
  \bibfield  {author} {\bibinfo {author} {\bibfnamefont {X.}~\bibnamefont
  {Wan}}, \bibinfo {author} {\bibfnamefont {A.~M.}\ \bibnamefont {Turner}},
  \bibinfo {author} {\bibfnamefont {A.}~\bibnamefont {Vishwanath}},\ and\
  \bibinfo {author} {\bibfnamefont {S.~Y.}\ \bibnamefont {Savrasov}},\
  }\bibfield  {title} {\bibinfo {title} {{Topological semimetal and Fermi-arc
  surface states in the electronic structure of pyrochlore iridates}},\ }\href
  {https://doi.org/10.1103/PhysRevB.83.205101} {\bibfield  {journal} {\bibinfo
  {journal} {Phys. Rev. B}\ }\textbf {\bibinfo {volume} {83}},\ \bibinfo
  {pages} {205101} (\bibinfo {year} {2011})}\BibitemShut {NoStop}%
\bibitem [{\citenamefont {Chang}\ \emph {et~al.}(2018)\citenamefont {Chang},
  \citenamefont {Wieder}, \citenamefont {Schindler}, \citenamefont {Sanchez},
  \citenamefont {Belopolski}, \citenamefont {Huang}, \citenamefont {Singh},
  \citenamefont {Wu}, \citenamefont {Chang}, \citenamefont {Neupert},
  \citenamefont {Xu}, \citenamefont {Lin},\ and\ \citenamefont
  {Hasan}}]{Kramers–Weyl-fermions}%
  \BibitemOpen
  \bibfield  {author} {\bibinfo {author} {\bibfnamefont {G.}~\bibnamefont
  {Chang}}, \bibinfo {author} {\bibfnamefont {B.~J.}\ \bibnamefont {Wieder}},
  \bibinfo {author} {\bibfnamefont {F.}~\bibnamefont {Schindler}}, \bibinfo
  {author} {\bibfnamefont {D.~S.}\ \bibnamefont {Sanchez}}, \bibinfo {author}
  {\bibfnamefont {I.}~\bibnamefont {Belopolski}}, \bibinfo {author}
  {\bibfnamefont {S.-M.}\ \bibnamefont {Huang}}, \bibinfo {author}
  {\bibfnamefont {B.}~\bibnamefont {Singh}}, \bibinfo {author} {\bibfnamefont
  {D.}~\bibnamefont {Wu}}, \bibinfo {author} {\bibfnamefont {T.-R.}\
  \bibnamefont {Chang}}, \bibinfo {author} {\bibfnamefont {T.}~\bibnamefont
  {Neupert}}, \bibinfo {author} {\bibfnamefont {S.-Y.}\ \bibnamefont {Xu}},
  \bibinfo {author} {\bibfnamefont {H.}~\bibnamefont {Lin}},\ and\ \bibinfo
  {author} {\bibfnamefont {M.~Z.}\ \bibnamefont {Hasan}},\ }\bibfield  {title}
  {\bibinfo {title} {Topological quantum properties of chiral crystals},\
  }\href {https://doi.org/10.1038/s41563-018-0169-3} {\bibfield  {journal}
  {\bibinfo  {journal} {Nature Materials}\ }\textbf {\bibinfo {volume} {17}},\
  \bibinfo {pages} {978} (\bibinfo {year} {2018})}\BibitemShut {NoStop}%
\bibitem [{\citenamefont {Bradlyn}\ \emph {et~al.}(2016)\citenamefont
  {Bradlyn}, \citenamefont {Cano}, \citenamefont {Wang}, \citenamefont
  {Vergniory}, \citenamefont {Felser}, \citenamefont {Cava},\ and\
  \citenamefont {Bernevig}}]{largetopocharge}%
  \BibitemOpen
  \bibfield  {author} {\bibinfo {author} {\bibfnamefont {B.}~\bibnamefont
  {Bradlyn}}, \bibinfo {author} {\bibfnamefont {J.}~\bibnamefont {Cano}},
  \bibinfo {author} {\bibfnamefont {Z.}~\bibnamefont {Wang}}, \bibinfo {author}
  {\bibfnamefont {M.~G.}\ \bibnamefont {Vergniory}}, \bibinfo {author}
  {\bibfnamefont {C.}~\bibnamefont {Felser}}, \bibinfo {author} {\bibfnamefont
  {R.~J.}\ \bibnamefont {Cava}},\ and\ \bibinfo {author} {\bibfnamefont
  {B.~A.}\ \bibnamefont {Bernevig}},\ }\bibfield  {title} {\bibinfo {title}
  {{Beyond Dirac and Weyl fermions: Unconventional quasiparticles in
  conventional crystals}},\ }\href {https://doi.org/10.1126/science.aaf5037}
  {\bibfield  {journal} {\bibinfo  {journal} {Science}\ }\textbf {\bibinfo
  {volume} {353}},\ \bibinfo {pages} {aaf5037} (\bibinfo {year}
  {2016})}\BibitemShut {NoStop}%
\bibitem [{\citenamefont {Chang}\ \emph {et~al.}(2017)\citenamefont {Chang},
  \citenamefont {Xu}, \citenamefont {Wieder}, \citenamefont {Sanchez},
  \citenamefont {Huang}, \citenamefont {Belopolski}, \citenamefont {Chang},
  \citenamefont {Zhang}, \citenamefont {Bansil}, \citenamefont {Lin},\ and\
  \citenamefont {Hasan}}]{longfermiarc1}%
  \BibitemOpen
  \bibfield  {author} {\bibinfo {author} {\bibfnamefont {G.}~\bibnamefont
  {Chang}}, \bibinfo {author} {\bibfnamefont {S.-Y.}\ \bibnamefont {Xu}},
  \bibinfo {author} {\bibfnamefont {B.~J.}\ \bibnamefont {Wieder}}, \bibinfo
  {author} {\bibfnamefont {D.~S.}\ \bibnamefont {Sanchez}}, \bibinfo {author}
  {\bibfnamefont {S.-M.}\ \bibnamefont {Huang}}, \bibinfo {author}
  {\bibfnamefont {I.}~\bibnamefont {Belopolski}}, \bibinfo {author}
  {\bibfnamefont {T.-R.}\ \bibnamefont {Chang}}, \bibinfo {author}
  {\bibfnamefont {S.}~\bibnamefont {Zhang}}, \bibinfo {author} {\bibfnamefont
  {A.}~\bibnamefont {Bansil}}, \bibinfo {author} {\bibfnamefont
  {H.}~\bibnamefont {Lin}},\ and\ \bibinfo {author} {\bibfnamefont {M.~Z.}\
  \bibnamefont {Hasan}},\ }\bibfield  {title} {\bibinfo {title}
  {{Unconventional Chiral Fermions and Large Topological Fermi Arcs in RhSi}},\
  }\href {https://doi.org/10.1103/PhysRevLett.119.206401} {\bibfield  {journal}
  {\bibinfo  {journal} {Phys. Rev. Lett.}\ }\textbf {\bibinfo {volume} {119}},\
  \bibinfo {pages} {206401} (\bibinfo {year} {2017})}\BibitemShut {NoStop}%
\bibitem [{\citenamefont {Tang}\ \emph {et~al.}(2017)\citenamefont {Tang},
  \citenamefont {Zhou},\ and\ \citenamefont {Zhang}}]{longfermiarc2}%
  \BibitemOpen
  \bibfield  {author} {\bibinfo {author} {\bibfnamefont {P.}~\bibnamefont
  {Tang}}, \bibinfo {author} {\bibfnamefont {Q.}~\bibnamefont {Zhou}},\ and\
  \bibinfo {author} {\bibfnamefont {S.-C.}\ \bibnamefont {Zhang}},\ }\bibfield
  {title} {\bibinfo {title} {{Multiple Types of Topological Fermions in
  Transition Metal Silicides}},\ }\href
  {https://doi.org/10.1103/PhysRevLett.119.206402} {\bibfield  {journal}
  {\bibinfo  {journal} {Phys. Rev. Lett.}\ }\textbf {\bibinfo {volume} {119}},\
  \bibinfo {pages} {206402} (\bibinfo {year} {2017})}\BibitemShut {NoStop}%
\bibitem [{\citenamefont {Zhong}\ \emph {et~al.}(2016)\citenamefont {Zhong},
  \citenamefont {Moore},\ and\ \citenamefont
  {Souza}}]{unusual-magnetotransport}%
  \BibitemOpen
  \bibfield  {author} {\bibinfo {author} {\bibfnamefont {S.}~\bibnamefont
  {Zhong}}, \bibinfo {author} {\bibfnamefont {J.~E.}\ \bibnamefont {Moore}},\
  and\ \bibinfo {author} {\bibfnamefont {I.}~\bibnamefont {Souza}},\ }\bibfield
   {title} {\bibinfo {title} {{Gyrotropic Magnetic Effect and the Magnetic
  Moment on the Fermi Surface}},\ }\href
  {https://doi.org/10.1103/PhysRevLett.116.077201} {\bibfield  {journal}
  {\bibinfo  {journal} {Phys. Rev. Lett.}\ }\textbf {\bibinfo {volume} {116}},\
  \bibinfo {pages} {077201} (\bibinfo {year} {2016})}\BibitemShut {NoStop}%
\bibitem [{\citenamefont {de~Juan}\ \emph {et~al.}(2017)\citenamefont
  {de~Juan}, \citenamefont {Grushin}, \citenamefont {Morimoto},\ and\
  \citenamefont {Moore}}]{quantized-circular-photogalvanic-effect}%
  \BibitemOpen
  \bibfield  {author} {\bibinfo {author} {\bibfnamefont {F.}~\bibnamefont
  {de~Juan}}, \bibinfo {author} {\bibfnamefont {A.~G.}\ \bibnamefont
  {Grushin}}, \bibinfo {author} {\bibfnamefont {T.}~\bibnamefont {Morimoto}},\
  and\ \bibinfo {author} {\bibfnamefont {J.~E.}\ \bibnamefont {Moore}},\
  }\bibfield  {title} {\bibinfo {title} {{Quantized circular photogalvanic
  effect in Weyl semimetals}},\ }\href {https://doi.org/10.1038/ncomms15995}
  {\bibfield  {journal} {\bibinfo  {journal} {Nat. Commun.}\ }\textbf {\bibinfo
  {volume} {8}},\ \bibinfo {pages} {15995} (\bibinfo {year}
  {2017})}\BibitemShut {NoStop}%
\bibitem [{\citenamefont {Lin}\ \emph {et~al.}(2022)\citenamefont {Lin},
  \citenamefont {Robredo}, \citenamefont {Schr\"oter}, \citenamefont {Felser},
  \citenamefont {Vergniory},\ and\ \citenamefont
  {Bradlyn}}]{Spin-momentum-locking1}%
  \BibitemOpen
  \bibfield  {author} {\bibinfo {author} {\bibfnamefont {M.}~\bibnamefont
  {Lin}}, \bibinfo {author} {\bibfnamefont {I.~n.}\ \bibnamefont {Robredo}},
  \bibinfo {author} {\bibfnamefont {N.~B.~M.}\ \bibnamefont {Schr\"oter}},
  \bibinfo {author} {\bibfnamefont {C.}~\bibnamefont {Felser}}, \bibinfo
  {author} {\bibfnamefont {M.~G.}\ \bibnamefont {Vergniory}},\ and\ \bibinfo
  {author} {\bibfnamefont {B.}~\bibnamefont {Bradlyn}},\ }\bibfield  {title}
  {\bibinfo {title} {{Spin-momentum locking from topological quantum chemistry:
  Applications to multifold fermions}},\ }\href
  {https://doi.org/10.1103/PhysRevB.106.245101} {\bibfield  {journal} {\bibinfo
   {journal} {Phys. Rev. B}\ }\textbf {\bibinfo {volume} {106}},\ \bibinfo
  {pages} {245101} (\bibinfo {year} {2022})}\BibitemShut {NoStop}%
\bibitem [{\citenamefont {Krieger}\ \emph {et~al.}(2024)\citenamefont
  {Krieger}, \citenamefont {Stolz}, \citenamefont {Robredo}, \citenamefont
  {Manna}, \citenamefont {McFarlane}, \citenamefont {Date}, \citenamefont
  {Pal}, \citenamefont {Yang}, \citenamefont {B.~Guedes}, \citenamefont {Dil},
  \citenamefont {Polley}, \citenamefont {Leandersson}, \citenamefont {Shekhar},
  \citenamefont {Borrmann}, \citenamefont {Yang}, \citenamefont {Lin},
  \citenamefont {Strocov}, \citenamefont {Caputo}, \citenamefont {Watson},
  \citenamefont {Kim}, \citenamefont {Cacho}, \citenamefont {Mazzola},
  \citenamefont {Fujii}, \citenamefont {Vobornik}, \citenamefont {Parkin},
  \citenamefont {Bradlyn}, \citenamefont {Felser}, \citenamefont {Vergniory},\
  and\ \citenamefont {Schr{\"o}ter}}]{Spin-momentum-locking2}%
  \BibitemOpen
  \bibfield  {author} {\bibinfo {author} {\bibfnamefont {J.~A.}\ \bibnamefont
  {Krieger}}, \bibinfo {author} {\bibfnamefont {S.}~\bibnamefont {Stolz}},
  \bibinfo {author} {\bibfnamefont {I.}~\bibnamefont {Robredo}}, \bibinfo
  {author} {\bibfnamefont {K.}~\bibnamefont {Manna}}, \bibinfo {author}
  {\bibfnamefont {E.~C.}\ \bibnamefont {McFarlane}}, \bibinfo {author}
  {\bibfnamefont {M.}~\bibnamefont {Date}}, \bibinfo {author} {\bibfnamefont
  {B.}~\bibnamefont {Pal}}, \bibinfo {author} {\bibfnamefont {J.}~\bibnamefont
  {Yang}}, \bibinfo {author} {\bibfnamefont {E.}~\bibnamefont {B.~Guedes}},
  \bibinfo {author} {\bibfnamefont {J.~H.}\ \bibnamefont {Dil}}, \bibinfo
  {author} {\bibfnamefont {C.~M.}\ \bibnamefont {Polley}}, \bibinfo {author}
  {\bibfnamefont {M.}~\bibnamefont {Leandersson}}, \bibinfo {author}
  {\bibfnamefont {C.}~\bibnamefont {Shekhar}}, \bibinfo {author} {\bibfnamefont
  {H.}~\bibnamefont {Borrmann}}, \bibinfo {author} {\bibfnamefont
  {Q.}~\bibnamefont {Yang}}, \bibinfo {author} {\bibfnamefont {M.}~\bibnamefont
  {Lin}}, \bibinfo {author} {\bibfnamefont {V.~N.}\ \bibnamefont {Strocov}},
  \bibinfo {author} {\bibfnamefont {M.}~\bibnamefont {Caputo}}, \bibinfo
  {author} {\bibfnamefont {M.~D.}\ \bibnamefont {Watson}}, \bibinfo {author}
  {\bibfnamefont {T.~K.}\ \bibnamefont {Kim}}, \bibinfo {author} {\bibfnamefont
  {C.}~\bibnamefont {Cacho}}, \bibinfo {author} {\bibfnamefont
  {F.}~\bibnamefont {Mazzola}}, \bibinfo {author} {\bibfnamefont
  {J.}~\bibnamefont {Fujii}}, \bibinfo {author} {\bibfnamefont
  {I.}~\bibnamefont {Vobornik}}, \bibinfo {author} {\bibfnamefont {S.~S.~P.}\
  \bibnamefont {Parkin}}, \bibinfo {author} {\bibfnamefont {B.}~\bibnamefont
  {Bradlyn}}, \bibinfo {author} {\bibfnamefont {C.}~\bibnamefont {Felser}},
  \bibinfo {author} {\bibfnamefont {M.~G.}\ \bibnamefont {Vergniory}},\ and\
  \bibinfo {author} {\bibfnamefont {N.~B.~M.}\ \bibnamefont {Schr{\"o}ter}},\
  }\bibfield  {title} {\bibinfo {title} {Weyl spin-momentum locking in a chiral
  topological semimetal},\ }\href {https://doi.org/10.1038/s41467-024-47976-0}
  {\bibfield  {journal} {\bibinfo  {journal} {Nature Communications}\ }\textbf
  {\bibinfo {volume} {15}},\ \bibinfo {pages} {3720} (\bibinfo {year}
  {2024})}\BibitemShut {NoStop}%
\bibitem [{\citenamefont {Yen}\ \emph {et~al.}(2024)\citenamefont {Yen},
  \citenamefont {Krieger}, \citenamefont {Yao}, \citenamefont {Robredo},
  \citenamefont {Manna}, \citenamefont {Yang}, \citenamefont {McFarlane},
  \citenamefont {Shekhar}, \citenamefont {Borrmann}, \citenamefont {Stolz},
  \citenamefont {Widmer}, \citenamefont {Gr{\"o}ning}, \citenamefont {Strocov},
  \citenamefont {Parkin}, \citenamefont {Felser}, \citenamefont {Vergniory},
  \citenamefont {Sch{\"u}ler},\ and\ \citenamefont
  {Schr{\"o}ter}}]{OAM-momentum-locking1}%
  \BibitemOpen
  \bibfield  {author} {\bibinfo {author} {\bibfnamefont {Y.}~\bibnamefont
  {Yen}}, \bibinfo {author} {\bibfnamefont {J.~A.}\ \bibnamefont {Krieger}},
  \bibinfo {author} {\bibfnamefont {M.}~\bibnamefont {Yao}}, \bibinfo {author}
  {\bibfnamefont {I.}~\bibnamefont {Robredo}}, \bibinfo {author} {\bibfnamefont
  {K.}~\bibnamefont {Manna}}, \bibinfo {author} {\bibfnamefont
  {Q.}~\bibnamefont {Yang}}, \bibinfo {author} {\bibfnamefont {E.~C.}\
  \bibnamefont {McFarlane}}, \bibinfo {author} {\bibfnamefont {C.}~\bibnamefont
  {Shekhar}}, \bibinfo {author} {\bibfnamefont {H.}~\bibnamefont {Borrmann}},
  \bibinfo {author} {\bibfnamefont {S.}~\bibnamefont {Stolz}}, \bibinfo
  {author} {\bibfnamefont {R.}~\bibnamefont {Widmer}}, \bibinfo {author}
  {\bibfnamefont {O.}~\bibnamefont {Gr{\"o}ning}}, \bibinfo {author}
  {\bibfnamefont {V.~N.}\ \bibnamefont {Strocov}}, \bibinfo {author}
  {\bibfnamefont {S.~S.~P.}\ \bibnamefont {Parkin}}, \bibinfo {author}
  {\bibfnamefont {C.}~\bibnamefont {Felser}}, \bibinfo {author} {\bibfnamefont
  {M.~G.}\ \bibnamefont {Vergniory}}, \bibinfo {author} {\bibfnamefont
  {M.}~\bibnamefont {Sch{\"u}ler}},\ and\ \bibinfo {author} {\bibfnamefont
  {N.~B.~M.}\ \bibnamefont {Schr{\"o}ter}},\ }\bibfield  {title} {\bibinfo
  {title} {Controllable orbital angular momentum monopoles in chiral
  topological semimetals},\ }\href {https://doi.org/10.1038/s41567-024-02655-1}
  {\bibfield  {journal} {\bibinfo  {journal} {Nature Physics}\ }\textbf
  {\bibinfo {volume} {20}},\ \bibinfo {pages} {1912} (\bibinfo {year}
  {2024})}\BibitemShut {NoStop}%
\bibitem [{\citenamefont {Liu}\ \emph {et~al.}(2017)\citenamefont {Liu},
  \citenamefont {Xu}, \citenamefont {Zhang},\ and\ \citenamefont
  {Duan}}]{app-waveguides-1}%
  \BibitemOpen
  \bibfield  {author} {\bibinfo {author} {\bibfnamefont {Y.}~\bibnamefont
  {Liu}}, \bibinfo {author} {\bibfnamefont {Y.}~\bibnamefont {Xu}}, \bibinfo
  {author} {\bibfnamefont {S.-C.}\ \bibnamefont {Zhang}},\ and\ \bibinfo
  {author} {\bibfnamefont {W.}~\bibnamefont {Duan}},\ }\bibfield  {title}
  {\bibinfo {title} {{Model for topological phononics and phonon diode}},\
  }\href {https://doi.org/10.1103/PhysRevB.96.064106} {\bibfield  {journal}
  {\bibinfo  {journal} {Phys. Rev. B}\ }\textbf {\bibinfo {volume} {96}},\
  \bibinfo {pages} {064106} (\bibinfo {year} {2017})}\BibitemShut {NoStop}%
\bibitem [{\citenamefont {Li}\ \emph {et~al.}(2005)\citenamefont {Li},
  \citenamefont {Wang}, \citenamefont {Wang},\ and\ \citenamefont
  {Zhang}}]{app-heat-2}%
  \BibitemOpen
  \bibfield  {author} {\bibinfo {author} {\bibfnamefont {B.}~\bibnamefont
  {Li}}, \bibinfo {author} {\bibfnamefont {J.}~\bibnamefont {Wang}}, \bibinfo
  {author} {\bibfnamefont {L.}~\bibnamefont {Wang}},\ and\ \bibinfo {author}
  {\bibfnamefont {G.}~\bibnamefont {Zhang}},\ }\bibfield  {title} {\bibinfo
  {title} {{Anomalous heat conduction and anomalous diffusion in nonlinear
  lattices, single walled nanotubes, and billiard gas channels}},\ }\href
  {https://doi.org/10.1063/1.1832791} {\bibfield  {journal} {\bibinfo
  {journal} {Chaos}\ }\textbf {\bibinfo {volume} {15}},\ \bibinfo {pages}
  {015121} (\bibinfo {year} {2005})}\BibitemShut {NoStop}%
\bibitem [{\citenamefont {Zhang}\ \emph {et~al.}(2016)\citenamefont {Zhang},
  \citenamefont {Guan}, \citenamefont {Jia}, \citenamefont {Liu}, \citenamefont
  {Wang}, \citenamefont {Li}, \citenamefont {Wang}, \citenamefont {Ma},
  \citenamefont {Xue}, \citenamefont {Zhang}, \citenamefont {Plummer},
  \citenamefont {Zhu},\ and\ \citenamefont {Guo}}]{en-super-2}%
  \BibitemOpen
  \bibfield  {author} {\bibinfo {author} {\bibfnamefont {S.}~\bibnamefont
  {Zhang}}, \bibinfo {author} {\bibfnamefont {J.}~\bibnamefont {Guan}},
  \bibinfo {author} {\bibfnamefont {X.}~\bibnamefont {Jia}}, \bibinfo {author}
  {\bibfnamefont {B.}~\bibnamefont {Liu}}, \bibinfo {author} {\bibfnamefont
  {W.}~\bibnamefont {Wang}}, \bibinfo {author} {\bibfnamefont {F.}~\bibnamefont
  {Li}}, \bibinfo {author} {\bibfnamefont {L.}~\bibnamefont {Wang}}, \bibinfo
  {author} {\bibfnamefont {X.}~\bibnamefont {Ma}}, \bibinfo {author}
  {\bibfnamefont {Q.}~\bibnamefont {Xue}}, \bibinfo {author} {\bibfnamefont
  {J.}~\bibnamefont {Zhang}}, \bibinfo {author} {\bibfnamefont {E.~W.}\
  \bibnamefont {Plummer}}, \bibinfo {author} {\bibfnamefont {X.}~\bibnamefont
  {Zhu}},\ and\ \bibinfo {author} {\bibfnamefont {J.}~\bibnamefont {Guo}},\
  }\bibfield  {title} {\bibinfo {title} {{Role of ${\mathrm{SrTiO}}_{3}$ phonon
  penetrating into thin FeSe films in the enhancement of superconductivity}},\
  }\href {https://doi.org/10.1103/PhysRevB.94.081116} {\bibfield  {journal}
  {\bibinfo  {journal} {Phys. Rev. B}\ }\textbf {\bibinfo {volume} {94}},\
  \bibinfo {pages} {081116} (\bibinfo {year} {2016})}\BibitemShut {NoStop}%
\bibitem [{\citenamefont {Singh}\ \emph {et~al.}(2018)\citenamefont {Singh},
  \citenamefont {Wu}, \citenamefont {Yue}, \citenamefont {Romero},\ and\
  \citenamefont {Soluyanov}}]{app-triple}%
  \BibitemOpen
  \bibfield  {author} {\bibinfo {author} {\bibfnamefont {S.}~\bibnamefont
  {Singh}}, \bibinfo {author} {\bibfnamefont {Q.~S.}\ \bibnamefont {Wu}},
  \bibinfo {author} {\bibfnamefont {C.}~\bibnamefont {Yue}}, \bibinfo {author}
  {\bibfnamefont {A.~H.}\ \bibnamefont {Romero}},\ and\ \bibinfo {author}
  {\bibfnamefont {A.~A.}\ \bibnamefont {Soluyanov}},\ }\bibfield  {title}
  {\bibinfo {title} {{Topological phonons and thermoelectricity in triple-point
  metals}},\ }\href {https://doi.org/10.1103/PhysRevMaterials.2.114204}
  {\bibfield  {journal} {\bibinfo  {journal} {Phys. Rev. Mater.}\ }\textbf
  {\bibinfo {volume} {2}},\ \bibinfo {pages} {114204} (\bibinfo {year}
  {2018})}\BibitemShut {NoStop}%
\bibitem [{\citenamefont {Li}\ \emph {et~al.}(2018)\citenamefont {Li},
  \citenamefont {Ma}, \citenamefont {Xie}, \citenamefont {Feng}, \citenamefont
  {Ullah}, \citenamefont {Li}, \citenamefont {Dong}, \citenamefont {Li},
  \citenamefont {Li},\ and\ \citenamefont {Chen}}]{chemical-1}%
  \BibitemOpen
  \bibfield  {author} {\bibinfo {author} {\bibfnamefont {J.}~\bibnamefont
  {Li}}, \bibinfo {author} {\bibfnamefont {H.}~\bibnamefont {Ma}}, \bibinfo
  {author} {\bibfnamefont {Q.}~\bibnamefont {Xie}}, \bibinfo {author}
  {\bibfnamefont {S.}~\bibnamefont {Feng}}, \bibinfo {author} {\bibfnamefont
  {S.}~\bibnamefont {Ullah}}, \bibinfo {author} {\bibfnamefont
  {R.}~\bibnamefont {Li}}, \bibinfo {author} {\bibfnamefont {J.}~\bibnamefont
  {Dong}}, \bibinfo {author} {\bibfnamefont {D.}~\bibnamefont {Li}}, \bibinfo
  {author} {\bibfnamefont {Y.}~\bibnamefont {Li}},\ and\ \bibinfo {author}
  {\bibfnamefont {X.-Q.}\ \bibnamefont {Chen}},\ }\bibfield  {title} {\bibinfo
  {title} {{Topological quantum catalyst: Dirac nodal line states and a
  potential electrocatalyst of hydrogen evolution in the TiSi family}},\ }\href
  {https://doi.org/10.1007/s40843-017-9178-4} {\bibfield  {journal} {\bibinfo
  {journal} {Sci. China Mater.}\ }\textbf {\bibinfo {volume} {61}},\ \bibinfo
  {pages} {23} (\bibinfo {year} {2018})}\BibitemShut {NoStop}%
\bibitem [{\citenamefont {Zhu}\ \emph {et~al.}(2022)\citenamefont {Zhu},
  \citenamefont {Wu}, \citenamefont {Zhao}, \citenamefont {Chen}, \citenamefont
  {Zhang},\ and\ \citenamefont {Yang}}]{chemical-2}%
  \BibitemOpen
  \bibfield  {author} {\bibinfo {author} {\bibfnamefont {J.}~\bibnamefont
  {Zhu}}, \bibinfo {author} {\bibfnamefont {W.}~\bibnamefont {Wu}}, \bibinfo
  {author} {\bibfnamefont {J.}~\bibnamefont {Zhao}}, \bibinfo {author}
  {\bibfnamefont {H.}~\bibnamefont {Chen}}, \bibinfo {author} {\bibfnamefont
  {L.}~\bibnamefont {Zhang}},\ and\ \bibinfo {author} {\bibfnamefont {S.~A.}\
  \bibnamefont {Yang}},\ }\bibfield  {title} {\bibinfo {title}
  {{Symmetry-enforced nodal chain phonons}},\ }\href
  {https://doi.org/10.1038/s41535-022-00461-7} {\bibfield  {journal} {\bibinfo
  {journal} {npj Quantum Mater.}\ }\textbf {\bibinfo {volume} {7}},\ \bibinfo
  {pages} {52} (\bibinfo {year} {2022})}\BibitemShut {NoStop}%
\bibitem [{\citenamefont {Hamada}\ \emph {et~al.}(2018)\citenamefont {Hamada},
  \citenamefont {Minamitani}, \citenamefont {Hirayama},\ and\ \citenamefont
  {Murakami}}]{phononHaaseffect}%
  \BibitemOpen
  \bibfield  {author} {\bibinfo {author} {\bibfnamefont {M.}~\bibnamefont
  {Hamada}}, \bibinfo {author} {\bibfnamefont {E.}~\bibnamefont {Minamitani}},
  \bibinfo {author} {\bibfnamefont {M.}~\bibnamefont {Hirayama}},\ and\
  \bibinfo {author} {\bibfnamefont {S.}~\bibnamefont {Murakami}},\ }\bibfield
  {title} {\bibinfo {title} {{Phonon Angular Momentum Induced by the
  Temperature Gradient}},\ }\href
  {https://doi.org/10.1103/PhysRevLett.121.175301} {\bibfield  {journal}
  {\bibinfo  {journal} {Phys. Rev. Lett.}\ }\textbf {\bibinfo {volume} {121}},\
  \bibinfo {pages} {175301} (\bibinfo {year} {2018})}\BibitemShut {NoStop}%
\bibitem [{\citenamefont {Kim}\ \emph {et~al.}(2023)\citenamefont {Kim},
  \citenamefont {Vetter}, \citenamefont {Yan}, \citenamefont {Yang},
  \citenamefont {Wang}, \citenamefont {Sun}, \citenamefont {Yang},
  \citenamefont {Comstock}, \citenamefont {Li}, \citenamefont {Zhou},
  \citenamefont {Zhang}, \citenamefont {You}, \citenamefont {Sun},\ and\
  \citenamefont {Liu}}]{spin-Seebeck-effect}%
  \BibitemOpen
  \bibfield  {author} {\bibinfo {author} {\bibfnamefont {K.}~\bibnamefont
  {Kim}}, \bibinfo {author} {\bibfnamefont {E.}~\bibnamefont {Vetter}},
  \bibinfo {author} {\bibfnamefont {L.}~\bibnamefont {Yan}}, \bibinfo {author}
  {\bibfnamefont {C.}~\bibnamefont {Yang}}, \bibinfo {author} {\bibfnamefont
  {Z.}~\bibnamefont {Wang}}, \bibinfo {author} {\bibfnamefont {R.}~\bibnamefont
  {Sun}}, \bibinfo {author} {\bibfnamefont {Y.}~\bibnamefont {Yang}}, \bibinfo
  {author} {\bibfnamefont {A.~H.}\ \bibnamefont {Comstock}}, \bibinfo {author}
  {\bibfnamefont {X.}~\bibnamefont {Li}}, \bibinfo {author} {\bibfnamefont
  {J.}~\bibnamefont {Zhou}}, \bibinfo {author} {\bibfnamefont {L.}~\bibnamefont
  {Zhang}}, \bibinfo {author} {\bibfnamefont {W.}~\bibnamefont {You}}, \bibinfo
  {author} {\bibfnamefont {D.}~\bibnamefont {Sun}},\ and\ \bibinfo {author}
  {\bibfnamefont {J.}~\bibnamefont {Liu}},\ }\bibfield  {title} {\bibinfo
  {title} {Chiral-phonon-activated spin seebeck effect},\ }\href
  {https://doi.org/10.1038/s41563-023-01473-9} {\bibfield  {journal} {\bibinfo
  {journal} {Nature Materials}\ }\textbf {\bibinfo {volume} {22}},\ \bibinfo
  {pages} {322} (\bibinfo {year} {2023})}\BibitemShut {NoStop}%
\bibitem [{\citenamefont {Strohm}\ \emph {et~al.}(2005)\citenamefont {Strohm},
  \citenamefont {Rikken},\ and\ \citenamefont {Wyder}}]{thermal-Hall-effect-1}%
  \BibitemOpen
  \bibfield  {author} {\bibinfo {author} {\bibfnamefont {C.}~\bibnamefont
  {Strohm}}, \bibinfo {author} {\bibfnamefont {G.~L. J.~A.}\ \bibnamefont
  {Rikken}},\ and\ \bibinfo {author} {\bibfnamefont {P.}~\bibnamefont
  {Wyder}},\ }\bibfield  {title} {\bibinfo {title} {{Phenomenological Evidence
  for the Phonon Hall Effect}},\ }\href
  {https://doi.org/10.1103/PhysRevLett.95.155901} {\bibfield  {journal}
  {\bibinfo  {journal} {Phys. Rev. Lett.}\ }\textbf {\bibinfo {volume} {95}},\
  \bibinfo {pages} {155901} (\bibinfo {year} {2005})}\BibitemShut {NoStop}%
\bibitem [{\citenamefont {Kagan}\ and\ \citenamefont
  {Maksimov}(2008)}]{thermal-Hall-effect-2}%
  \BibitemOpen
  \bibfield  {author} {\bibinfo {author} {\bibfnamefont {Y.}~\bibnamefont
  {Kagan}}\ and\ \bibinfo {author} {\bibfnamefont {L.~A.}\ \bibnamefont
  {Maksimov}},\ }\bibfield  {title} {\bibinfo {title} {{Anomalous Hall Effect
  for the Phonon Heat Conductivity in Paramagnetic Dielectrics}},\ }\href
  {https://doi.org/10.1103/PhysRevLett.100.145902} {\bibfield  {journal}
  {\bibinfo  {journal} {Phys. Rev. Lett.}\ }\textbf {\bibinfo {volume} {100}},\
  \bibinfo {pages} {145902} (\bibinfo {year} {2008})}\BibitemShut {NoStop}%
\bibitem [{\citenamefont {Grissonnanche}\ \emph {et~al.}(2020)\citenamefont
  {Grissonnanche}, \citenamefont {Th{\'e}riault}, \citenamefont {Gourgout},
  \citenamefont {Boulanger}, \citenamefont {Lefran{\c{c}}ois}, \citenamefont
  {Ataei}, \citenamefont {Lalibert{\'e}}, \citenamefont {Dion}, \citenamefont
  {Zhou}, \citenamefont {Pyon}, \citenamefont {Takayama}, \citenamefont
  {Takagi}, \citenamefont {Doiron-Leyraud},\ and\ \citenamefont
  {Taillefer}}]{thermal-Hall-effect-3}%
  \BibitemOpen
  \bibfield  {author} {\bibinfo {author} {\bibfnamefont {G.}~\bibnamefont
  {Grissonnanche}}, \bibinfo {author} {\bibfnamefont {S.}~\bibnamefont
  {Th{\'e}riault}}, \bibinfo {author} {\bibfnamefont {A.}~\bibnamefont
  {Gourgout}}, \bibinfo {author} {\bibfnamefont {M.-E.}\ \bibnamefont
  {Boulanger}}, \bibinfo {author} {\bibfnamefont {E.}~\bibnamefont
  {Lefran{\c{c}}ois}}, \bibinfo {author} {\bibfnamefont {A.}~\bibnamefont
  {Ataei}}, \bibinfo {author} {\bibfnamefont {F.}~\bibnamefont
  {Lalibert{\'e}}}, \bibinfo {author} {\bibfnamefont {M.}~\bibnamefont {Dion}},
  \bibinfo {author} {\bibfnamefont {J.-S.}\ \bibnamefont {Zhou}}, \bibinfo
  {author} {\bibfnamefont {S.}~\bibnamefont {Pyon}}, \bibinfo {author}
  {\bibfnamefont {T.}~\bibnamefont {Takayama}}, \bibinfo {author}
  {\bibfnamefont {H.}~\bibnamefont {Takagi}}, \bibinfo {author} {\bibfnamefont
  {N.}~\bibnamefont {Doiron-Leyraud}},\ and\ \bibinfo {author} {\bibfnamefont
  {L.}~\bibnamefont {Taillefer}},\ }\bibfield  {title} {\bibinfo {title}
  {Chiral phonons in the pseudogap phase of cuprates},\ }\href
  {https://doi.org/10.1038/s41567-020-0965-y} {\bibfield  {journal} {\bibinfo
  {journal} {Nature Physics}\ }\textbf {\bibinfo {volume} {16}},\ \bibinfo
  {pages} {1108} (\bibinfo {year} {2020})}\BibitemShut {NoStop}%
\bibitem [{\citenamefont {Ren}\ \emph {et~al.}(2021)\citenamefont {Ren},
  \citenamefont {Xiao}, \citenamefont {Saparov},\ and\ \citenamefont
  {Niu}}]{PMM1}%
  \BibitemOpen
  \bibfield  {author} {\bibinfo {author} {\bibfnamefont {Y.}~\bibnamefont
  {Ren}}, \bibinfo {author} {\bibfnamefont {C.}~\bibnamefont {Xiao}}, \bibinfo
  {author} {\bibfnamefont {D.}~\bibnamefont {Saparov}},\ and\ \bibinfo {author}
  {\bibfnamefont {Q.}~\bibnamefont {Niu}},\ }\bibfield  {title} {\bibinfo
  {title} {{Phonon Magnetic Moment from Electronic Topological
  Magnetization}},\ }\href {https://doi.org/10.1103/PhysRevLett.127.186403}
  {\bibfield  {journal} {\bibinfo  {journal} {Phys. Rev. Lett.}\ }\textbf
  {\bibinfo {volume} {127}},\ \bibinfo {pages} {186403} (\bibinfo {year}
  {2021})}\BibitemShut {NoStop}%
\bibitem [{\citenamefont {Hernandez}\ \emph {et~al.}(2023)\citenamefont
  {Hernandez}, \citenamefont {Baydin}, \citenamefont {Chaudhary}, \citenamefont
  {Tay}, \citenamefont {Katayama}, \citenamefont {Takeda}, \citenamefont
  {Nojiri}, \citenamefont {Okazaki}, \citenamefont {Rappl}, \citenamefont
  {Abramof}, \citenamefont {Rodriguez-Vega}, \citenamefont {Fiete},\ and\
  \citenamefont {Kono}}]{PMM2}%
  \BibitemOpen
  \bibfield  {author} {\bibinfo {author} {\bibfnamefont {F.~G.}\ \bibnamefont
  {Hernandez}}, \bibinfo {author} {\bibfnamefont {A.}~\bibnamefont {Baydin}},
  \bibinfo {author} {\bibfnamefont {S.}~\bibnamefont {Chaudhary}}, \bibinfo
  {author} {\bibfnamefont {F.}~\bibnamefont {Tay}}, \bibinfo {author}
  {\bibfnamefont {I.}~\bibnamefont {Katayama}}, \bibinfo {author}
  {\bibfnamefont {J.}~\bibnamefont {Takeda}}, \bibinfo {author} {\bibfnamefont
  {H.}~\bibnamefont {Nojiri}}, \bibinfo {author} {\bibfnamefont {A.~K.}\
  \bibnamefont {Okazaki}}, \bibinfo {author} {\bibfnamefont {P.~H.}\
  \bibnamefont {Rappl}}, \bibinfo {author} {\bibfnamefont {E.}~\bibnamefont
  {Abramof}}, \bibinfo {author} {\bibfnamefont {M.}~\bibnamefont
  {Rodriguez-Vega}}, \bibinfo {author} {\bibfnamefont {G.~A.}\ \bibnamefont
  {Fiete}},\ and\ \bibinfo {author} {\bibfnamefont {J.}~\bibnamefont {Kono}},\
  }\bibfield  {title} {\bibinfo {title} {Observation of interplay between
  phonon chirality and electronic band topology},\ }\href
  {https://doi.org/10.1126/sciadv.adj4074} {\bibfield  {journal} {\bibinfo
  {journal} {Science Advances}\ }\textbf {\bibinfo {volume} {9}},\ \bibinfo
  {pages} {eadj4074} (\bibinfo {year} {2023})}\BibitemShut {NoStop}%
\bibitem [{\citenamefont {Luo}\ \emph {et~al.}(2023)\citenamefont {Luo},
  \citenamefont {Lin}, \citenamefont {Zhang}, \citenamefont {Chen},
  \citenamefont {Blackert}, \citenamefont {Xu}, \citenamefont {Yakobson},\ and\
  \citenamefont {Zhu}}]{PMM3}%
  \BibitemOpen
  \bibfield  {author} {\bibinfo {author} {\bibfnamefont {J.}~\bibnamefont
  {Luo}}, \bibinfo {author} {\bibfnamefont {T.}~\bibnamefont {Lin}}, \bibinfo
  {author} {\bibfnamefont {J.}~\bibnamefont {Zhang}}, \bibinfo {author}
  {\bibfnamefont {X.}~\bibnamefont {Chen}}, \bibinfo {author} {\bibfnamefont
  {E.~R.}\ \bibnamefont {Blackert}}, \bibinfo {author} {\bibfnamefont
  {R.}~\bibnamefont {Xu}}, \bibinfo {author} {\bibfnamefont {B.~I.}\
  \bibnamefont {Yakobson}},\ and\ \bibinfo {author} {\bibfnamefont
  {H.}~\bibnamefont {Zhu}},\ }\bibfield  {title} {\bibinfo {title} {Large
  effective magnetic fields from chiral phonons in rare-earth halides},\ }\href
  {https://doi.org/10.1126/science.adi9601} {\bibfield  {journal} {\bibinfo
  {journal} {Science}\ }\textbf {\bibinfo {volume} {382}},\ \bibinfo {pages}
  {698} (\bibinfo {year} {2023})}\BibitemShut {NoStop}%
\bibitem [{\citenamefont {Wang}\ \emph
  {et~al.}(2025{\natexlab{a}})\citenamefont {Wang}, \citenamefont {Liu},
  \citenamefont {Sun}, \citenamefont {Wang}, \citenamefont {Murakami},
  \citenamefont {Zhang}, \citenamefont {Zhang},\ and\ \citenamefont
  {Xing}}]{PMM4}%
  \BibitemOpen
  \bibfield  {author} {\bibinfo {author} {\bibfnamefont {F.}~\bibnamefont
  {Wang}}, \bibinfo {author} {\bibfnamefont {X.}~\bibnamefont {Liu}}, \bibinfo
  {author} {\bibfnamefont {H.}~\bibnamefont {Sun}}, \bibinfo {author}
  {\bibfnamefont {H.}~\bibnamefont {Wang}}, \bibinfo {author} {\bibfnamefont
  {S.}~\bibnamefont {Murakami}}, \bibinfo {author} {\bibfnamefont
  {L.}~\bibnamefont {Zhang}}, \bibinfo {author} {\bibfnamefont
  {H.}~\bibnamefont {Zhang}},\ and\ \bibinfo {author} {\bibfnamefont
  {D.}~\bibnamefont {Xing}},\ }\bibfield  {title} {\bibinfo {title} {{Ab Initio
  Theory of Phonon Magnetic Moment Induced by Electron-Phonon Coupling in
  Magnetic Materials}},\ }\href {https://doi.org/10.1103/tpjd-dh1m} {\bibfield
  {journal} {\bibinfo  {journal} {Phys. Rev. Lett.}\ }\textbf {\bibinfo
  {volume} {135}},\ \bibinfo {pages} {256701} (\bibinfo {year}
  {2025}{\natexlab{a}})}\BibitemShut {NoStop}%
\bibitem [{\citenamefont {Zhang}\ and\ \citenamefont {Niu}(2014)}]{Lifa-L-1}%
  \BibitemOpen
  \bibfield  {author} {\bibinfo {author} {\bibfnamefont {L.}~\bibnamefont
  {Zhang}}\ and\ \bibinfo {author} {\bibfnamefont {Q.}~\bibnamefont {Niu}},\
  }\bibfield  {title} {\bibinfo {title} {{Angular Momentum of Phonons and the
  Einstein--de Haas Effect}},\ }\href
  {https://doi.org/10.1103/PhysRevLett.112.085503} {\bibfield  {journal}
  {\bibinfo  {journal} {Phys. Rev. Lett.}\ }\textbf {\bibinfo {volume} {112}},\
  \bibinfo {pages} {085503} (\bibinfo {year} {2014})}\BibitemShut {NoStop}%
\bibitem [{\citenamefont {Zhang}\ and\ \citenamefont {Niu}(2015)}]{Lifa-L-2}%
  \BibitemOpen
  \bibfield  {author} {\bibinfo {author} {\bibfnamefont {L.}~\bibnamefont
  {Zhang}}\ and\ \bibinfo {author} {\bibfnamefont {Q.}~\bibnamefont {Niu}},\
  }\bibfield  {title} {\bibinfo {title} {{Chiral Phonons at High-Symmetry
  Points in Monolayer Hexagonal Lattices}},\ }\href
  {https://doi.org/10.1103/PhysRevLett.115.115502} {\bibfield  {journal}
  {\bibinfo  {journal} {Phys. Rev. Lett.}\ }\textbf {\bibinfo {volume} {115}},\
  \bibinfo {pages} {115502} (\bibinfo {year} {2015})}\BibitemShut {NoStop}%
\bibitem [{\citenamefont {Juraschek}\ \emph {et~al.}(2025)\citenamefont
  {Juraschek}, \citenamefont {Geilhufe}, \citenamefont {Zhu}, \citenamefont
  {Basini}, \citenamefont {Baum}, \citenamefont {Baydin}, \citenamefont
  {Chaudhary}, \citenamefont {Fechner}, \citenamefont {Flebus}, \citenamefont
  {Grissonnanche}, \citenamefont {Kirilyuk}, \citenamefont {Lemeshko},
  \citenamefont {Maehrlein}, \citenamefont {Mignolet}, \citenamefont
  {Murakami}, \citenamefont {Niu}, \citenamefont {Nowak}, \citenamefont
  {Romao}, \citenamefont {Rostami}, \citenamefont {Satoh}, \citenamefont
  {Spaldin}, \citenamefont {Ueda},\ and\ \citenamefont
  {Zhang}}]{Chiralphonons}%
  \BibitemOpen
  \bibfield  {author} {\bibinfo {author} {\bibfnamefont {D.~M.}\ \bibnamefont
  {Juraschek}}, \bibinfo {author} {\bibfnamefont {R.~M.}\ \bibnamefont
  {Geilhufe}}, \bibinfo {author} {\bibfnamefont {H.}~\bibnamefont {Zhu}},
  \bibinfo {author} {\bibfnamefont {M.}~\bibnamefont {Basini}}, \bibinfo
  {author} {\bibfnamefont {P.}~\bibnamefont {Baum}}, \bibinfo {author}
  {\bibfnamefont {A.}~\bibnamefont {Baydin}}, \bibinfo {author} {\bibfnamefont
  {S.}~\bibnamefont {Chaudhary}}, \bibinfo {author} {\bibfnamefont
  {M.}~\bibnamefont {Fechner}}, \bibinfo {author} {\bibfnamefont
  {B.}~\bibnamefont {Flebus}}, \bibinfo {author} {\bibfnamefont
  {G.}~\bibnamefont {Grissonnanche}}, \bibinfo {author} {\bibfnamefont {A.~I.}\
  \bibnamefont {Kirilyuk}}, \bibinfo {author} {\bibfnamefont {M.}~\bibnamefont
  {Lemeshko}}, \bibinfo {author} {\bibfnamefont {S.~F.}\ \bibnamefont
  {Maehrlein}}, \bibinfo {author} {\bibfnamefont {M.}~\bibnamefont {Mignolet}},
  \bibinfo {author} {\bibfnamefont {S.}~\bibnamefont {Murakami}}, \bibinfo
  {author} {\bibfnamefont {Q.}~\bibnamefont {Niu}}, \bibinfo {author}
  {\bibfnamefont {U.}~\bibnamefont {Nowak}}, \bibinfo {author} {\bibfnamefont
  {C.~P.}\ \bibnamefont {Romao}}, \bibinfo {author} {\bibfnamefont
  {H.}~\bibnamefont {Rostami}}, \bibinfo {author} {\bibfnamefont
  {T.}~\bibnamefont {Satoh}}, \bibinfo {author} {\bibfnamefont {N.~A.}\
  \bibnamefont {Spaldin}}, \bibinfo {author} {\bibfnamefont {H.}~\bibnamefont
  {Ueda}},\ and\ \bibinfo {author} {\bibfnamefont {L.}~\bibnamefont {Zhang}},\
  }\bibfield  {title} {\bibinfo {title} {Chiral phonons},\ }\href
  {https://doi.org/10.1038/s41567-025-03001-9} {\bibfield  {journal} {\bibinfo
  {journal} {Nature Physics}\ }\textbf {\bibinfo {volume} {21}},\ \bibinfo
  {pages} {1532} (\bibinfo {year} {2025})}\BibitemShut {NoStop}%
\bibitem [{\citenamefont {Zhu}\ \emph {et~al.}(2018)\citenamefont {Zhu},
  \citenamefont {Yi}, \citenamefont {Li}, \citenamefont {Xiao}, \citenamefont
  {Zhang}, \citenamefont {Yang}, \citenamefont {Kaindl}, \citenamefont {Li},
  \citenamefont {Wang},\ and\ \citenamefont {Zhang}}]{chiral-obser}%
  \BibitemOpen
  \bibfield  {author} {\bibinfo {author} {\bibfnamefont {H.}~\bibnamefont
  {Zhu}}, \bibinfo {author} {\bibfnamefont {J.}~\bibnamefont {Yi}}, \bibinfo
  {author} {\bibfnamefont {M.-Y.}\ \bibnamefont {Li}}, \bibinfo {author}
  {\bibfnamefont {J.}~\bibnamefont {Xiao}}, \bibinfo {author} {\bibfnamefont
  {L.}~\bibnamefont {Zhang}}, \bibinfo {author} {\bibfnamefont {C.-W.}\
  \bibnamefont {Yang}}, \bibinfo {author} {\bibfnamefont {R.~A.}\ \bibnamefont
  {Kaindl}}, \bibinfo {author} {\bibfnamefont {L.-J.}\ \bibnamefont {Li}},
  \bibinfo {author} {\bibfnamefont {Y.}~\bibnamefont {Wang}},\ and\ \bibinfo
  {author} {\bibfnamefont {X.}~\bibnamefont {Zhang}},\ }\bibfield  {title}
  {\bibinfo {title} {{Observation of chiral phonons}},\ }\href
  {https://doi.org/10.1126/science.aar2711} {\bibfield  {journal} {\bibinfo
  {journal} {Science}\ }\textbf {\bibinfo {volume} {359}},\ \bibinfo {pages}
  {579} (\bibinfo {year} {2018})}\BibitemShut {NoStop}%
\bibitem [{\citenamefont {Ueda}\ \emph {et~al.}(2023)\citenamefont {Ueda},
  \citenamefont {Garcia-Fernandez}, \citenamefont {Agrestini}, \citenamefont
  {Romao}, \citenamefont {van~den Brink}, \citenamefont {Spaldin},
  \citenamefont {Zhou},\ and\ \citenamefont {Staub}}]{SiO2}%
  \BibitemOpen
  \bibfield  {author} {\bibinfo {author} {\bibfnamefont {H.}~\bibnamefont
  {Ueda}}, \bibinfo {author} {\bibfnamefont {M.}~\bibnamefont
  {Garcia-Fernandez}}, \bibinfo {author} {\bibfnamefont {S.}~\bibnamefont
  {Agrestini}}, \bibinfo {author} {\bibfnamefont {C.~P.}\ \bibnamefont
  {Romao}}, \bibinfo {author} {\bibfnamefont {J.}~\bibnamefont {van~den
  Brink}}, \bibinfo {author} {\bibfnamefont {N.~A.}\ \bibnamefont {Spaldin}},
  \bibinfo {author} {\bibfnamefont {K.-J.}\ \bibnamefont {Zhou}},\ and\
  \bibinfo {author} {\bibfnamefont {U.}~\bibnamefont {Staub}},\ }\bibfield
  {title} {\bibinfo {title} {{Chiral phonons in quartz probed by X-rays}},\
  }\href {https://doi.org/10.1038/s41586-023-06016-5} {\bibfield  {journal}
  {\bibinfo  {journal} {Nature}\ }\textbf {\bibinfo {volume} {618}},\ \bibinfo
  {pages} {946} (\bibinfo {year} {2023})}\BibitemShut {NoStop}%
\bibitem [{\citenamefont {Zhang}\ \emph {et~al.}(2023)\citenamefont {Zhang},
  \citenamefont {Huang}, \citenamefont {Pan}, \citenamefont {Du}, \citenamefont
  {Zhang},\ and\ \citenamefont {Murakami}}]{Weyl-Chiral}%
  \BibitemOpen
  \bibfield  {author} {\bibinfo {author} {\bibfnamefont {T.}~\bibnamefont
  {Zhang}}, \bibinfo {author} {\bibfnamefont {Z.}~\bibnamefont {Huang}},
  \bibinfo {author} {\bibfnamefont {Z.}~\bibnamefont {Pan}}, \bibinfo {author}
  {\bibfnamefont {L.}~\bibnamefont {Du}}, \bibinfo {author} {\bibfnamefont
  {G.}~\bibnamefont {Zhang}},\ and\ \bibinfo {author} {\bibfnamefont
  {S.}~\bibnamefont {Murakami}},\ }\bibfield  {title} {\bibinfo {title} {{Weyl
  Phonons in Chiral Crystals}},\ }\href
  {https://doi.org/10.1021/acs.nanolett.3c02132} {\bibfield  {journal}
  {\bibinfo  {journal} {Nano Lett.}\ }\textbf {\bibinfo {volume} {23}},\
  \bibinfo {pages} {7561} (\bibinfo {year} {2023})}\BibitemShut {NoStop}%
\bibitem [{\citenamefont {Ishito}\ \emph
  {et~al.}(2023{\natexlab{a}})\citenamefont {Ishito}, \citenamefont {Mao},
  \citenamefont {Kousaka}, \citenamefont {Togawa}, \citenamefont {Iwasaki},
  \citenamefont {Zhang}, \citenamefont {Murakami}, \citenamefont {Kishine},\
  and\ \citenamefont {Satoh}}]{chrialHgS}%
  \BibitemOpen
  \bibfield  {author} {\bibinfo {author} {\bibfnamefont {K.}~\bibnamefont
  {Ishito}}, \bibinfo {author} {\bibfnamefont {H.}~\bibnamefont {Mao}},
  \bibinfo {author} {\bibfnamefont {Y.}~\bibnamefont {Kousaka}}, \bibinfo
  {author} {\bibfnamefont {Y.}~\bibnamefont {Togawa}}, \bibinfo {author}
  {\bibfnamefont {S.}~\bibnamefont {Iwasaki}}, \bibinfo {author} {\bibfnamefont
  {T.}~\bibnamefont {Zhang}}, \bibinfo {author} {\bibfnamefont
  {S.}~\bibnamefont {Murakami}}, \bibinfo {author} {\bibfnamefont {J.-i.}\
  \bibnamefont {Kishine}},\ and\ \bibinfo {author} {\bibfnamefont
  {T.}~\bibnamefont {Satoh}},\ }\bibfield  {title} {\bibinfo {title} {{Truly
  chiral phonons in $\alpha-$HgS}},\ }\href
  {https://doi.org/10.1038/s41567-022-01790-x} {\bibfield  {journal} {\bibinfo
  {journal} {Nature Physics}\ }\textbf {\bibinfo {volume} {19}},\ \bibinfo
  {pages} {35} (\bibinfo {year} {2023}{\natexlab{a}})}\BibitemShut {NoStop}%
\bibitem [{\citenamefont {Ishito}\ \emph
  {et~al.}(2023{\natexlab{b}})\citenamefont {Ishito}, \citenamefont {Mao},
  \citenamefont {Kobayashi}, \citenamefont {Kousaka}, \citenamefont {Togawa},
  \citenamefont {Kusunose}, \citenamefont {Kishine},\ and\ \citenamefont
  {Satoh}}]{Chirality}%
  \BibitemOpen
  \bibfield  {author} {\bibinfo {author} {\bibfnamefont {K.}~\bibnamefont
  {Ishito}}, \bibinfo {author} {\bibfnamefont {H.}~\bibnamefont {Mao}},
  \bibinfo {author} {\bibfnamefont {K.}~\bibnamefont {Kobayashi}}, \bibinfo
  {author} {\bibfnamefont {Y.}~\bibnamefont {Kousaka}}, \bibinfo {author}
  {\bibfnamefont {Y.}~\bibnamefont {Togawa}}, \bibinfo {author} {\bibfnamefont
  {H.}~\bibnamefont {Kusunose}}, \bibinfo {author} {\bibfnamefont {J.-i.}\
  \bibnamefont {Kishine}},\ and\ \bibinfo {author} {\bibfnamefont
  {T.}~\bibnamefont {Satoh}},\ }\bibfield  {title} {\bibinfo {title} {{Chiral
  phonons: circularly polarized Raman spectroscopy and ab initio calculations
  in a chiral crystal tellurium}},\ }\href
  {https://doi.org/https://doi.org/10.1002/chir.23544} {\bibfield  {journal}
  {\bibinfo  {journal} {Chirality}\ }\textbf {\bibinfo {volume} {35}},\
  \bibinfo {pages} {338} (\bibinfo {year} {2023}{\natexlab{b}})}\BibitemShut
  {NoStop}%
\bibitem [{\citenamefont {Ueda}\ \emph {et~al.}(2025)\citenamefont {Ueda},
  \citenamefont {Nag}, \citenamefont {Romao}, \citenamefont
  {García-Fernández}, \citenamefont {Zhou},\ and\ \citenamefont
  {Staub}}]{Ueda2025}%
  \BibitemOpen
  \bibfield  {author} {\bibinfo {author} {\bibfnamefont {H.}~\bibnamefont
  {Ueda}}, \bibinfo {author} {\bibfnamefont {A.}~\bibnamefont {Nag}}, \bibinfo
  {author} {\bibfnamefont {C.~P.}\ \bibnamefont {Romao}}, \bibinfo {author}
  {\bibfnamefont {M.}~\bibnamefont {García-Fernández}}, \bibinfo {author}
  {\bibfnamefont {K.-J.}\ \bibnamefont {Zhou}},\ and\ \bibinfo {author}
  {\bibfnamefont {U.}~\bibnamefont {Staub}},\ }\bibfield  {title} {\bibinfo
  {title} {Chiral phonons in polar {LiNbO3}},\ }\href
  {https://doi.org/10.1038/s41467-025-66911-5} {\bibfield  {journal} {\bibinfo
  {journal} {Nature Communications}\ }\textbf {\bibinfo {volume} {17}},\
  \bibinfo {pages} {212} (\bibinfo {year} {2025})}\BibitemShut {NoStop}%
\bibitem [{\citenamefont {Zhang}\ \emph
  {et~al.}(2025{\natexlab{a}})\citenamefont {Zhang}, \citenamefont {Murakami},\
  and\ \citenamefont {Miao}}]{connection1}%
  \BibitemOpen
  \bibfield  {author} {\bibinfo {author} {\bibfnamefont {T.}~\bibnamefont
  {Zhang}}, \bibinfo {author} {\bibfnamefont {S.}~\bibnamefont {Murakami}},\
  and\ \bibinfo {author} {\bibfnamefont {H.}~\bibnamefont {Miao}},\ }\bibfield
  {title} {\bibinfo {title} {{Weyl phonons: the connection of topology and
  chirality}},\ }\href {https://doi.org/10.1038/s41467-025-58913-0} {\bibfield
  {journal} {\bibinfo  {journal} {Nature Communications}\ }\textbf {\bibinfo
  {volume} {16}},\ \bibinfo {pages} {3560} (\bibinfo {year}
  {2025}{\natexlab{a}})}\BibitemShut {NoStop}%
\bibitem [{\citenamefont {Zhang}\ \emph
  {et~al.}(2025{\natexlab{b}})\citenamefont {Zhang}, \citenamefont {Liu},
  \citenamefont {Miao},\ and\ \citenamefont {Murakami}}]{connection2}%
  \BibitemOpen
  \bibfield  {author} {\bibinfo {author} {\bibfnamefont {T.}~\bibnamefont
  {Zhang}}, \bibinfo {author} {\bibfnamefont {Y.}~\bibnamefont {Liu}}, \bibinfo
  {author} {\bibfnamefont {H.}~\bibnamefont {Miao}},\ and\ \bibinfo {author}
  {\bibfnamefont {S.}~\bibnamefont {Murakami}},\ }\href
  {https://arxiv.org/abs/2505.06179} {\bibinfo {title} {{Advances in Phonons:
  From Band Topology to Phonon Chirality}}} (\bibinfo {year}
  {2025}{\natexlab{b}}),\ \Eprint {https://arxiv.org/abs/2505.06179}
  {arXiv:2505.06179} \BibitemShut {NoStop}%
\bibitem [{\citenamefont {Zhang}\ \emph {et~al.}(2024)\citenamefont {Zhang},
  \citenamefont {Luo},\ and\ \citenamefont {Zhang}}]{charge-density-wave}%
  \BibitemOpen
  \bibfield  {author} {\bibinfo {author} {\bibfnamefont {S.}~\bibnamefont
  {Zhang}}, \bibinfo {author} {\bibfnamefont {K.}~\bibnamefont {Luo}},\ and\
  \bibinfo {author} {\bibfnamefont {T.}~\bibnamefont {Zhang}},\ }\bibfield
  {title} {\bibinfo {title} {Understanding chiral charge-density wave by frozen
  chiral phonon},\ }\href {https://doi.org/10.1038/s41524-024-01440-1}
  {\bibfield  {journal} {\bibinfo  {journal} {npj Computational Materials}\
  }\textbf {\bibinfo {volume} {10}},\ \bibinfo {pages} {264} (\bibinfo {year}
  {2024})}\BibitemShut {NoStop}%
\bibitem [{\citenamefont {Tauchert}\ \emph {et~al.}(2022)\citenamefont
  {Tauchert}, \citenamefont {Volkov}, \citenamefont {Ehberger}, \citenamefont
  {Kazenwadel}, \citenamefont {Evers}, \citenamefont {Lange}, \citenamefont
  {Donges}, \citenamefont {Book}, \citenamefont {Kreuzpaintner}, \citenamefont
  {Nowak},\ and\ \citenamefont {Baum}}]{ultrafast-demagnetization}%
  \BibitemOpen
  \bibfield  {author} {\bibinfo {author} {\bibfnamefont {S.~R.}\ \bibnamefont
  {Tauchert}}, \bibinfo {author} {\bibfnamefont {M.}~\bibnamefont {Volkov}},
  \bibinfo {author} {\bibfnamefont {D.}~\bibnamefont {Ehberger}}, \bibinfo
  {author} {\bibfnamefont {D.}~\bibnamefont {Kazenwadel}}, \bibinfo {author}
  {\bibfnamefont {M.}~\bibnamefont {Evers}}, \bibinfo {author} {\bibfnamefont
  {H.}~\bibnamefont {Lange}}, \bibinfo {author} {\bibfnamefont
  {A.}~\bibnamefont {Donges}}, \bibinfo {author} {\bibfnamefont
  {A.}~\bibnamefont {Book}}, \bibinfo {author} {\bibfnamefont {W.}~\bibnamefont
  {Kreuzpaintner}}, \bibinfo {author} {\bibfnamefont {U.}~\bibnamefont
  {Nowak}},\ and\ \bibinfo {author} {\bibfnamefont {P.}~\bibnamefont {Baum}},\
  }\bibfield  {title} {\bibinfo {title} {Polarized phonons carry angular
  momentum in ultrafast demagnetization},\ }\href
  {https://doi.org/10.1038/s41586-021-04306-4} {\bibfield  {journal} {\bibinfo
  {journal} {Nature}\ }\textbf {\bibinfo {volume} {602}},\ \bibinfo {pages}
  {73} (\bibinfo {year} {2022})}\BibitemShut {NoStop}%
\bibitem [{\citenamefont {Mulliken}(1933)}]{Mulliken1939}%
  \BibitemOpen
  \bibfield  {author} {\bibinfo {author} {\bibfnamefont {R.~S.}\ \bibnamefont
  {Mulliken}},\ }\bibfield  {title} {\bibinfo {title} {{Electronic Structures
  of Polyatomic Molecules and Valence. IV. Electronic States, Quantum Theory of
  the Double Bond}},\ }\href {https://doi.org/10.1103/PhysRev.43.279}
  {\bibfield  {journal} {\bibinfo  {journal} {Phys. Rev.}\ }\textbf {\bibinfo
  {volume} {43}},\ \bibinfo {pages} {279} (\bibinfo {year} {1933})}\BibitemShut
  {NoStop}%
\bibitem [{\citenamefont {Hellenbrandt}(2004)}]{ICSD}%
  \BibitemOpen
  \bibfield  {author} {\bibinfo {author} {\bibfnamefont {M.}~\bibnamefont
  {Hellenbrandt}},\ }\bibfield  {title} {\bibinfo {title} {{The inorganic
  crystal structure database (ICSD)—present and future}},\ }\href
  {https://doi.org/10.1080/08893110410001664882} {\bibfield  {journal}
  {\bibinfo  {journal} {Crystallogr. Rev.}\ }\textbf {\bibinfo {volume} {10}},\
  \bibinfo {pages} {17} (\bibinfo {year} {2004})}\BibitemShut {NoStop}%
\bibitem [{pho()}]{phonopyDB}%
  \BibitemOpen
  \href@noop {} {}\bibinfo {howpublished}
  {\url{https://github.com/atztogo/phonondb}}\BibitemShut {NoStop}%
\bibitem [{\citenamefont {Tang}\ and\ \citenamefont
  {Wan}(2024{\natexlab{b}})}]{NNL2}%
  \BibitemOpen
  \bibfield  {author} {\bibinfo {author} {\bibfnamefont {F.}~\bibnamefont
  {Tang}}\ and\ \bibinfo {author} {\bibfnamefont {X.}~\bibnamefont {Wan}},\
  }\bibfield  {title} {\bibinfo {title} {{Group-theoretical study of band nodes
  and the emanating nodal structures in crystalline materials}},\ }\href
  {https://doi.org/10.1007/s44214-024-00060-6} {\bibfield  {journal} {\bibinfo
  {journal} {Quantum Front.}\ }\textbf {\bibinfo {volume} {3}},\ \bibinfo
  {pages} {14} (\bibinfo {year} {2024}{\natexlab{b}})}\BibitemShut {NoStop}%
\bibitem [{\citenamefont {Guo}\ \emph {et~al.}(2022)\citenamefont {Guo},
  \citenamefont {Hu}, \citenamefont {Putzke}, \citenamefont {Diaz},
  \citenamefont {Huang}, \citenamefont {Manna}, \citenamefont {Fan},
  \citenamefont {Shekhar}, \citenamefont {Sun}, \citenamefont {Felser},
  \citenamefont {Liu}, \citenamefont {Bernevig},\ and\ \citenamefont
  {Moll}}]{NNL3}%
  \BibitemOpen
  \bibfield  {author} {\bibinfo {author} {\bibfnamefont {C.}~\bibnamefont
  {Guo}}, \bibinfo {author} {\bibfnamefont {L.}~\bibnamefont {Hu}}, \bibinfo
  {author} {\bibfnamefont {C.}~\bibnamefont {Putzke}}, \bibinfo {author}
  {\bibfnamefont {J.}~\bibnamefont {Diaz}}, \bibinfo {author} {\bibfnamefont
  {X.}~\bibnamefont {Huang}}, \bibinfo {author} {\bibfnamefont
  {K.}~\bibnamefont {Manna}}, \bibinfo {author} {\bibfnamefont {F.-R.}\
  \bibnamefont {Fan}}, \bibinfo {author} {\bibfnamefont {C.}~\bibnamefont
  {Shekhar}}, \bibinfo {author} {\bibfnamefont {Y.}~\bibnamefont {Sun}},
  \bibinfo {author} {\bibfnamefont {C.}~\bibnamefont {Felser}}, \bibinfo
  {author} {\bibfnamefont {C.}~\bibnamefont {Liu}}, \bibinfo {author}
  {\bibfnamefont {B.~A.}\ \bibnamefont {Bernevig}},\ and\ \bibinfo {author}
  {\bibfnamefont {P.~J.~W.}\ \bibnamefont {Moll}},\ }\bibfield  {title}
  {\bibinfo {title} {{Quasi-symmetry-protected topology in a semi-metal}},\
  }\href {https://doi.org/10.1038/s41567-022-01604-0} {\bibfield  {journal}
  {\bibinfo  {journal} {Nat. Phys.}\ }\textbf {\bibinfo {volume} {18}},\
  \bibinfo {pages} {813} (\bibinfo {year} {2022})}\BibitemShut {NoStop}%
\bibitem [{\citenamefont {Hu}\ \emph {et~al.}(2023)\citenamefont {Hu},
  \citenamefont {Guo}, \citenamefont {Sun}, \citenamefont {Felser},
  \citenamefont {Elcoro}, \citenamefont {Moll}, \citenamefont {Liu},\ and\
  \citenamefont {Bernevig}}]{NNL4}%
  \BibitemOpen
  \bibfield  {author} {\bibinfo {author} {\bibfnamefont {L.-H.}\ \bibnamefont
  {Hu}}, \bibinfo {author} {\bibfnamefont {C.}~\bibnamefont {Guo}}, \bibinfo
  {author} {\bibfnamefont {Y.}~\bibnamefont {Sun}}, \bibinfo {author}
  {\bibfnamefont {C.}~\bibnamefont {Felser}}, \bibinfo {author} {\bibfnamefont
  {L.}~\bibnamefont {Elcoro}}, \bibinfo {author} {\bibfnamefont {P.~J.~W.}\
  \bibnamefont {Moll}}, \bibinfo {author} {\bibfnamefont {C.-X.}\ \bibnamefont
  {Liu}},\ and\ \bibinfo {author} {\bibfnamefont {B.~A.}\ \bibnamefont
  {Bernevig}},\ }\bibfield  {title} {\bibinfo {title} {{Hierarchy of
  quasisymmetries and degeneracies in the CoSi family of chiral crystal
  materials}},\ }\href {https://doi.org/10.1103/PhysRevB.107.125145} {\bibfield
   {journal} {\bibinfo  {journal} {Phys. Rev. B}\ }\textbf {\bibinfo {volume}
  {107}},\ \bibinfo {pages} {125145} (\bibinfo {year} {2023})}\BibitemShut
  {NoStop}%
\bibitem [{\citenamefont {Fan}\ \emph {et~al.}(2023)\citenamefont {Fan},
  \citenamefont {Wan},\ and\ \citenamefont {Tang}}]{NNL5}%
  \BibitemOpen
  \bibfield  {author} {\bibinfo {author} {\bibfnamefont {D.}~\bibnamefont
  {Fan}}, \bibinfo {author} {\bibfnamefont {X.}~\bibnamefont {Wan}},\ and\
  \bibinfo {author} {\bibfnamefont {F.}~\bibnamefont {Tang}},\ }\bibfield
  {title} {\bibinfo {title} {{Catalog of maximally charged Weyl points hosting
  nearly emanating nodal lines in phonon spectra}},\ }\href
  {https://doi.org/10.1103/PhysRevB.108.104110} {\bibfield  {journal} {\bibinfo
   {journal} {Phys. Rev. B}\ }\textbf {\bibinfo {volume} {108}},\ \bibinfo
  {pages} {104110} (\bibinfo {year} {2023})}\BibitemShut {NoStop}%
\bibitem [{\citenamefont {Li}\ \emph {et~al.}(2024)\citenamefont {Li},
  \citenamefont {Zhang}, \citenamefont {Liu},\ and\ \citenamefont
  {Liu}}]{NNL6}%
  \BibitemOpen
  \bibfield  {author} {\bibinfo {author} {\bibfnamefont {J.}~\bibnamefont
  {Li}}, \bibinfo {author} {\bibfnamefont {A.}~\bibnamefont {Zhang}}, \bibinfo
  {author} {\bibfnamefont {Y.}~\bibnamefont {Liu}},\ and\ \bibinfo {author}
  {\bibfnamefont {Q.}~\bibnamefont {Liu}},\ }\bibfield  {title} {\bibinfo
  {title} {{Group Theory on Quasisymmetry and Protected Near Degeneracy}},\
  }\href {https://doi.org/10.1103/PhysRevLett.133.026402} {\bibfield  {journal}
  {\bibinfo  {journal} {Phys. Rev. Lett.}\ }\textbf {\bibinfo {volume} {133}},\
  \bibinfo {pages} {026402} (\bibinfo {year} {2024})}\BibitemShut {NoStop}%
\bibitem [{\citenamefont {Liu}\ \emph {et~al.}(2024)\citenamefont {Liu},
  \citenamefont {Liu}, \citenamefont {Li}, \citenamefont {Wu},\ and\
  \citenamefont {Liu}}]{NNL7}%
  \BibitemOpen
  \bibfield  {author} {\bibinfo {author} {\bibfnamefont {L.}~\bibnamefont
  {Liu}}, \bibinfo {author} {\bibfnamefont {Y.}~\bibnamefont {Liu}}, \bibinfo
  {author} {\bibfnamefont {J.}~\bibnamefont {Li}}, \bibinfo {author}
  {\bibfnamefont {H.}~\bibnamefont {Wu}},\ and\ \bibinfo {author}
  {\bibfnamefont {Q.}~\bibnamefont {Liu}},\ }\bibfield  {title} {\bibinfo
  {title} {{Quantum spin Hall effect protected by spin U(1) quasisymmetry}},\
  }\href {https://doi.org/10.1103/PhysRevB.110.L161104} {\bibfield  {journal}
  {\bibinfo  {journal} {Phys. Rev. B}\ }\textbf {\bibinfo {volume} {110}},\
  \bibinfo {pages} {L161104} (\bibinfo {year} {2024})}\BibitemShut {NoStop}%
\bibitem [{\citenamefont {Yang}\ \emph {et~al.}(2025)\citenamefont {Yang},
  \citenamefont {Xiao}, \citenamefont {Mao}, \citenamefont {Li}, \citenamefont
  {Wang}, \citenamefont {Deng}, \citenamefont {Tang}, \citenamefont {Song},
  \citenamefont {Li}, \citenamefont {Yuan}, \citenamefont {Shi},\ and\
  \citenamefont {Xu}}]{phononAMtexture1}%
  \BibitemOpen
  \bibfield  {author} {\bibinfo {author} {\bibfnamefont {Y.}~\bibnamefont
  {Yang}}, \bibinfo {author} {\bibfnamefont {Z.}~\bibnamefont {Xiao}}, \bibinfo
  {author} {\bibfnamefont {Y.}~\bibnamefont {Mao}}, \bibinfo {author}
  {\bibfnamefont {Z.}~\bibnamefont {Li}}, \bibinfo {author} {\bibfnamefont
  {Z.}~\bibnamefont {Wang}}, \bibinfo {author} {\bibfnamefont {T.}~\bibnamefont
  {Deng}}, \bibinfo {author} {\bibfnamefont {Y.}~\bibnamefont {Tang}}, \bibinfo
  {author} {\bibfnamefont {Z.-D.}\ \bibnamefont {Song}}, \bibinfo {author}
  {\bibfnamefont {Y.}~\bibnamefont {Li}}, \bibinfo {author} {\bibfnamefont
  {H.}~\bibnamefont {Yuan}}, \bibinfo {author} {\bibfnamefont {M.}~\bibnamefont
  {Shi}},\ and\ \bibinfo {author} {\bibfnamefont {Y.}~\bibnamefont {Xu}},\
  }\href {https://arxiv.org/abs/2506.13721} {\bibinfo {title} {{Catalogue of
  chiral phonon materials}}} (\bibinfo {year} {2025}),\ \Eprint
  {https://arxiv.org/abs/2506.13721} {arXiv:2506.13721} \BibitemShut {NoStop}%
\bibitem [{\citenamefont {Liu}(2022)}]{phononAMtexture2}%
  \BibitemOpen
  \bibfield  {author} {\bibinfo {author} {\bibfnamefont {C.-X.}\ \bibnamefont
  {Liu}},\ }\bibfield  {title} {\bibinfo {title} {{Probing Nieh-Yan anomaly
  through phonon dynamics in the Kramers-Weyl semimetals of chiral crystals}},\
  }\href {https://doi.org/10.1103/PhysRevB.106.115102} {\bibfield  {journal}
  {\bibinfo  {journal} {Phys. Rev. B}\ }\textbf {\bibinfo {volume} {106}},\
  \bibinfo {pages} {115102} (\bibinfo {year} {2022})}\BibitemShut {NoStop}%
\bibitem [{\citenamefont {Wang}\ \emph
  {et~al.}(2025{\natexlab{b}})\citenamefont {Wang}, \citenamefont {Xu},
  \citenamefont {Liu}, \citenamefont {Wang}, \citenamefont {Zhang},\ and\
  \citenamefont {Zhang}}]{phononAMtexture3}%
  \BibitemOpen
  \bibfield  {author} {\bibinfo {author} {\bibfnamefont {F.}~\bibnamefont
  {Wang}}, \bibinfo {author} {\bibfnamefont {J.}~\bibnamefont {Xu}}, \bibinfo
  {author} {\bibfnamefont {X.}~\bibnamefont {Liu}}, \bibinfo {author}
  {\bibfnamefont {H.}~\bibnamefont {Wang}}, \bibinfo {author} {\bibfnamefont
  {L.}~\bibnamefont {Zhang}},\ and\ \bibinfo {author} {\bibfnamefont
  {H.}~\bibnamefont {Zhang}},\ }\href {https://arxiv.org/abs/2512.07518}
  {\bibinfo {title} {Alteraxial phonons in collinear magnets}} (\bibinfo {year}
  {2025}{\natexlab{b}}),\ \Eprint {https://arxiv.org/abs/2512.07518}
  {arXiv:2512.07518} \BibitemShut {NoStop}%
\bibitem [{\citenamefont {Bendin}\ \emph {et~al.}(2025)\citenamefont {Bendin},
  \citenamefont {Mook}, \citenamefont {Mertig},\ and\ \citenamefont
  {Neumann}}]{phononAMtexture4}%
  \BibitemOpen
  \bibfield  {author} {\bibinfo {author} {\bibfnamefont {H.}~\bibnamefont
  {Bendin}}, \bibinfo {author} {\bibfnamefont {A.}~\bibnamefont {Mook}},
  \bibinfo {author} {\bibfnamefont {I.}~\bibnamefont {Mertig}},\ and\ \bibinfo
  {author} {\bibfnamefont {R.~R.}\ \bibnamefont {Neumann}},\ }\href
  {https://arxiv.org/abs/2511.08357} {\bibinfo {title} {{D-Wave Phonon Angular
  Momentum Texture in Altermagnets by Magnon-Phonon-Hybridization}}} (\bibinfo
  {year} {2025}),\ \Eprint {https://arxiv.org/abs/2511.08357}
  {arXiv:2511.08357} \BibitemShut {NoStop}%
\bibitem [{\citenamefont {Juraschek}\ \emph {et~al.}(2017)\citenamefont
  {Juraschek}, \citenamefont {Fechner}, \citenamefont {Balatsky},\ and\
  \citenamefont {Spaldin}}]{orbitalMM1}%
  \BibitemOpen
  \bibfield  {author} {\bibinfo {author} {\bibfnamefont {D.~M.}\ \bibnamefont
  {Juraschek}}, \bibinfo {author} {\bibfnamefont {M.}~\bibnamefont {Fechner}},
  \bibinfo {author} {\bibfnamefont {A.~V.}\ \bibnamefont {Balatsky}},\ and\
  \bibinfo {author} {\bibfnamefont {N.~A.}\ \bibnamefont {Spaldin}},\
  }\bibfield  {title} {\bibinfo {title} {Dynamical multiferroicity},\ }\href
  {https://doi.org/10.1103/PhysRevMaterials.1.014401} {\bibfield  {journal}
  {\bibinfo  {journal} {Phys. Rev. Mater.}\ }\textbf {\bibinfo {volume} {1}},\
  \bibinfo {pages} {014401} (\bibinfo {year} {2017})}\BibitemShut {NoStop}%
\bibitem [{\citenamefont {Juraschek}\ and\ \citenamefont
  {Spaldin}(2019)}]{orbitalMM2}%
  \BibitemOpen
  \bibfield  {author} {\bibinfo {author} {\bibfnamefont {D.~M.}\ \bibnamefont
  {Juraschek}}\ and\ \bibinfo {author} {\bibfnamefont {N.~A.}\ \bibnamefont
  {Spaldin}},\ }\bibfield  {title} {\bibinfo {title} {Orbital magnetic moments
  of phonons},\ }\href {https://doi.org/10.1103/PhysRevMaterials.3.064405}
  {\bibfield  {journal} {\bibinfo  {journal} {Phys. Rev. Mater.}\ }\textbf
  {\bibinfo {volume} {3}},\ \bibinfo {pages} {064405} (\bibinfo {year}
  {2019})}\BibitemShut {NoStop}%
\end{thebibliography}
\end{document}